\documentclass[a4paper,10pt]{cpc-hepnp}
\usepackage{CJK,upgreek,fancyhdr}
\usepackage{multicol}
\usepackage{graphicx}
\usepackage{subfigure}
\usepackage{color}
\graphicspath{{figures/}{fig/}}
\usepackage{amsmath}
\usepackage{bm}
\usepackage{slashed}
\usepackage{epsfig}
\usepackage{amsfonts}
\usepackage{epstopdf}
\usepackage{hyperref}
\usepackage{booktabs}
\usepackage{multirow}
\usepackage{float}
\usepackage{makecell}
\usepackage{graphicx}
\allowdisplaybreaks
\begin{document}

\title{NLO effects for $\Omega_{QQQ}$ Baryons in QCD Sum Rules}

\author{ Ren-Hua Wu$^{1}$,\email{renhuawu@pku.edu.cn}
Yu-Sheng Zuo$^{1}$,\email{1801210125@pku.edu.cn}
Ce Meng$^{1}$,\email{mengce75@pku.edu.cn}
Yan-Qing Ma$^{1,2,3}$,\email{yqma@pku.edu.cn}
Kuang-Ta Chao$^{1,2,3}$,\email{ktchao@pku.edu.cn}}
\maketitle

\address{
$^{1}$School of Physics and State Key Laboratory of Nuclear Physics and Technology, Peking University, Beijing 100871, China\\
$^{2}$Center for High Energy Physics,
Peking University, Beijing 100871, China\\
$^{3}$Collaborative Innovation Center of Quantum Matter,
Beijing 100871, China
}%


\begin{abstract}
{\bf We study the triply heavy baryons $\Omega_{QQQ}$  $(Q=c, b)$ in the QCD sum rules by performing the first calculation of the next-to-leading order (NLO) contribution to the perturbative QCD part of the correlation functions.  Compared with the leading order (LO) result, the NLO contribution is found to be very important to the $\Omega_{QQQ}$. This is because the NLO not only results in a large correction, but also  reduces the parameter dependence, making the Borel platform more distinct, especially for the $\Omega_{bbb}$ in the $\overline{\rm{MS}}$ scheme, where the platform appears only at NLO but not at LO.
Particularly, owing to the inclusion of the NLO contribution, the  renormalization schemes ($\bar {MS}$  and On-Shell) dependence and the scale dependence are significantly reduced.
Consequently, after including the NLO contribution to the perturbative part in the QCD sum rules, the masses are estimated to be $4.53^{+0.26}_{-0.11}$ GeV for $\Omega_{ccc}$ and $14.27^{+0.33}_{-0.32}$  GeV for $\Omega_{bbb}$, where the results are obtained at $\mu=M_B$ with errors including those from the variation of the renormalization scale $\mu$ in the range $(0.8-1.2) M_B$. A careful study of the $\mu$ dependence in a wide range is further performed, which shows that the LO results are very sensitive to the choice of $\mu$ whereas the NLO results are considerably better. In addition to the $\mu=M_B$ result, a more stable value, (4.75-4.80) GeV, for the $\Omega_{ccc}$ mass is found in the range of $\mu=(1.2-2.0) M_B$ which should be viewed as a more relevant prediction in our NLO approach because of $\mu$ dependence.}

\end{abstract}

\maketitle


\section{Introduction}

In recent years, a large number of new hadronic states containing heavy quarks (charm quark $c$ or bottom quark $b$) have been observed in hadron colliders and $e^+e^-$ colliders\cite{Tanabashi:2018oca}. For example, the tetraquark states, pentaquark states, and baryons which contain two heavy quarks \cite{Chen:2016qju,Liu:2019zoy,Brambilla:2019esw}. These findings have opened up a new stage for the study of hadron physics and QCD.
It is well known that the light flavor baryons are composed of three light quarks ($u, d, s$), and all kinds of light flavor baryons from $\Delta^{++}(uuu)$ to $\Omega^- (sss)$ have been observed for many decades.
For the heavy baryons, the singly charmed and bottom baryon (containing one heavy quark $c$ or $b$ and two light quarks) have already formed a subfamily within heavy hadrons in recent years. Remarkably, the doubly charmed baryon $\Xi^{++}_{cc}(ccu)$, which contains two charm quarks and one light quark, was discovered in 2017 \cite{Aaij:2017ueg}, and more of its kind are expected to be found soon. The discovery of the doubly charmed baryon is an important event, and it may indicate that the whole family of baryons with all flavors $(u,d,s,c,b)$ may be found in the not far future. Here, the last member of the baryon family, i.e., the triply heavy baryons, which are composed of three heavy quarks $c$ or $b$, are yet to be discovered.

The $\Omega_{QQQ}$ baryons are made of three identical heavy quarks ($Q=c, b$). Because of their special properties in the baryon family, the $\Omega_{QQQ}$ baryons have been studied extensively for their productions\cite{Saleev:1999ti,GomshiNobary:2003sf,GomshiNobary:2004mq,GomshiNobary:2005ur,Chen:2011mb,He:2014tga,Baranov:2004er}, weak decays\cite{Bjorken:1985ei,Flynn:2011gf,Wang:2018utj}, masses\cite{Briceno:2012wt,Meinel:2012qz,Padmanath:2013zfa,Namekawa:2013vu,Brown:2014ena,Alexandrou:2014sha,Can:2015exa,Zhang:2009re,Aliev:2014lxa,Wang:2020avt,Hasenfratz:1980ka,Vijande:2004at,Migura:2006ep,Roberts:2007ni,Martynenko:2007je,Bernotas:2008bu,Vijande:2015faa,Thakkar:2016sog,Chen:2016spr,Weng:2018mmf,Yang:2019lsg,Liu:2019vtx,Radin:2014yna,Qin:2019hgk,Yin:2019bxe,Gutierrez-Guerrero:2019uwa,Wei:2015gsa,Wei:2016jyk} and so on. Particularly, the masses of $\Omega_{QQQ}$ baryons have been evaluated using various approaches, including lattice QCD \cite{Briceno:2012wt,Meinel:2012qz,Padmanath:2013zfa,Namekawa:2013vu,Brown:2014ena,Alexandrou:2014sha,Can:2015exa}, QCD sum rules~\cite{Zhang:2009re,Aliev:2014lxa,Wang:2020avt}, potential models~\cite{Hasenfratz:1980ka,Vijande:2004at,Migura:2006ep,Roberts:2007ni,Martynenko:2007je,Bernotas:2008bu,Vijande:2015faa,Thakkar:2016sog,Chen:2016spr,Weng:2018mmf,Yang:2019lsg,Liu:2019vtx}, the Fadeev equation \cite{Radin:2014yna,Qin:2019hgk,Yin:2019bxe,Gutierrez-Guerrero:2019uwa}, and  the Regge trajectories \cite{Wei:2015gsa,Wei:2016jyk}. The predicted values for the masses of $\Omega_{QQQ}$ baryons are listed in Table~\ref{tab:3c-3b-DiffModels}. As shown, the predicted masses are in a wide range, and different approaches produce different results. Therefore, it is still necessary to examine the $\Omega_{QQQ}$ baryons.

In this paper, we present the study of $\Omega_{QQQ}$ baryon using the QCD sum rules~\cite{Shifman:1978bx,Shifman:1978by,Reinders:1984sr} approach, which is known to be a powerful tool to evaluate hadron properties \cite{Colangelo:2000dp,Narison:2010wb,Narison:2014wqa,Albuquerque:2018jkn}. In this approach, one starts at short distances and then moves to long distances using the operator product expansion (OPE) for a given quark current's correlation function in QCD. The first term in the OPE is the dimensionless (d=0) identity operator, which represents the perturbative  QCD part, and then, as power corrections, the higher dimensional operators (d=4,6,...) with vacuum condensations emerge, which represent the nonperturbative (confinement) contribution.
Principally, because of the power suppression ($\frac{\Lambda _{QCD}^2}{-q^2}\ll 1$) at small distances (large $(-q^2)$), the high dimensional operator contributions should decrease as the power increases, one may only need to consider a few important terms in the OPE. Moreover, for the first (Identity) term in the OPE, i.e., the perturbative QCD term, one needs to consider not only the leading-order (LO) but also, at least, the next-to-leading order (NLO) $O(\alpha_s)$ contributions ($\alpha_s$ being the strong coupling constant), because the latter may lead to substantial corrections. Additionally, the $O(\alpha_s)$ corrections to the coefficients of the power-suppressed condensation terms may also need to be considered.
Practically, for the heavy $\Omega_{QQQ}$ system, the most important contributions in the OPE are the perturbative term $C_1$ and gluon condensation term $C_{GG} \langle g_s^2 \hat{G}\hat{G} \rangle$, where $C_1$ and $C_{GG}$ can be calculated perturbatively. As a good approximation, it may be necessary to consider the LO of $C_1$, which is the most important contribution, the NLO of $C_1$, which gives large corrections, and the LO of $C_{GG}$, which is the same order of magnitude as the former, but neglects other contributions. In fact, the importance of including the NLO contribution of $C_1$ has been emphasized in many studies, e.g., for
the proton (uud)~\cite{Ovchinnikov:1991mu,Groote:2008hz}, the singly heavy baryon~\cite{Groote:2008dx}, and the doubly heavy baryon $\Xi_{cc}^{++}$~\cite{Wang:2017qvg}. Our previous work~\cite{Wang:2017qvg} analyzed the NLO effect thoroughly for the doubly heavy baryon $\Xi_{cc}^{++}$ and determined that the NLO correction is sizable for $\Xi_{cc}^{++}$, and can't be ignored in QCD sum rules. Accordingly, we expect that the NLO correction also produces a sizable contribution to triply heavy baryons ($QQQ$).
Presently, no work has been conducted at the NLO level for triply heavy baryons. However, for some leading order (LO) results of the $\Omega_{QQQ}$ \cite{Zhang:2009re,Aliev:2014lxa,Wang:2020avt}, there are significant differences among different studies, as shown in the preceding paragraph. To reduce the uncertainties at the LO, it may be necessary to perform the NLO calculations in QCD sum rules. With the inclusion of the NLO contribution, the result should be substantially improved. Moreover, it is worthwhile to emphasize that there are significant differences between the fully heavy baryons and other baryons that contain light quarks. For the former, the most important nonperturbative contribution comes from the gluon condensation $\left \langle GG  \right \rangle$, whereas for the latter, the light quark condensation $\left \langle \bar{q}q  \right \rangle$ makes the important contribution. This point will be embodied in our calculation for $\Omega_{QQQ}$ in the QCD sum rules.

The rest of the paper is organized as follows. In Sec.~\ref{sec:QCD sum rules}, the sum rules for the calculation of the mass of $\Omega_{QQQ}$ are presented. In Sec.~\ref{sec:Calculation of C1 and CGG}, we introduce the methods and procedures for the calculation of the coefficients $C_1$ and $C_{GG}$. The phenomenology results and discussions are presented in Sec.~\ref{sec:Phenomenology}.


\section{QCD Sum Rules}\label{sec:QCD sum rules}

For the S-wave triply heavy quark $(QQQ)$ system, owing to Fermi-Dirac statistics~\cite{Ioffe:1981kw,Wang:2020avt}, there only exists the $ J^P=\frac{3}{2}^+$ baryon ground state $\Omega_{QQQ}$ ($Q=c,\,b$), with the corresponding current $J_{\mu}$~\cite{Ioffe:1981kw,Zhang:2009re,Wang:2020avt},
\begin{align}\label{eq:R+-meson}
J_{\mu} = \epsilon^{abc}(Q^T_a \hat{C}\gamma^\mu Q_b)\, Q_c\,,
\end{align}
where $a$, $b$, and $c$ denote the color indices, and $\hat{C}$ is the charge-conjugation matrix. Because there is no QQQ ground state with $ J=\frac{1}{2}$, it is not necessary to worry about the pollution from the $ J=\frac{1}{2}$ state in the analysis.

In the QCD sum rules, we begin with the two-point correlation function
\begin{align}\label{eq:corrFun}
\Pi_{\mu \nu}(q^2) &= i \int {d^4 x e^{iq \cdot x} \langle \Omega|T[J_\mu (x) \bar{J}_\nu(0)] |\Omega\rangle}, \
\end{align}
where $\Omega$ represents the QCD vacuum.
According to the lorentz covariance, the matrix element of the current can be written as
\begin{align}\label{eq:current-baryon}
\langle \Omega|J_{\mu}(0) |H(q,s)\rangle =\sqrt{\lambda} U_\mu(q,s),
\end{align}
where $H(q,s)$ denotes the ground state baryon with mass $M_H$, momentum $q$, and spin $s$; $\lambda$ is the pole residue for $H$, and $U_\mu(q,s)$ is the corresponding Rarita-Schwinger spinor, which satisfies the relation
\begin{align}\label{eq:spinorSum}
\sum_s U_\mu(q,s)\bar{U}_\nu(q,s)=(\slashed{q} +M_H)\left(-g_{\mu \nu} +\frac{\gamma_\mu \gamma_\nu}{3} + \frac{2 q_\mu q_\nu}{3 M_H^2}-\frac{q_\mu \gamma_\nu-\gamma_\mu q_\nu}{3M_H} \right).
\end{align}
Correlation function $\Pi_{\mu \nu}$ can then be written in the form
\begin{align}\label{eq:VAcorrFun}
\begin{split}
\Pi_{\mu \nu}(q^2) =& M  \left(-g_{\mu \nu} \Pi_1 (q^2)+\frac{\gamma_\mu \gamma_\nu}{3} \Pi_2 (q^2) + \frac{2 q_\mu q_\nu}{3 M^2}\Pi_3 (q^2) -\frac{q_\mu \gamma_\nu-\gamma_\mu q_\nu}{3M} \Pi_4 (q^2)\right) +\\
&\slashed{q} \left(-g_{\mu \nu} \Pi_5 (q^2)+\frac{\gamma_\mu \gamma_\nu}{3} \Pi_6 (q^2) + \frac{2 q_\mu q_\nu}{3 M^2}\Pi_7 (q^2) -\frac{q_\mu \gamma_\nu-\gamma_\mu q_\nu}{3M} \Pi_8 (q^2)\right)\,.
\end{split}
\end{align}
In this paper, we choose $\Pi_1(q^2)$ for the calculation to obtain mass $M_H$ of the ground state in the QCD sum rules. For convenience, in the following we use $\Pi(q^2)$ to denote $\Pi_1(q^2)$.

However, correlation function $\Pi(q^2)$ can be related to the phenomenological spectrum by the  K\"{a}ll\'{e}n-Lehmann representation
\begin{align}\label{eq:K-LSR}
\Pi(q^2) &= - \int {ds \frac{\rho(s)}{q^2-s+i\epsilon}}\,,
\end{align}
where $\rho(s)$ is the spectrum density, which contains information about all resonances and the continuum. Taking the narrow resonance approximation for the physical ground state, we may assume the spectrum density to consist of a single pole and a continuum spectrum, where all excited states are included in the continuum spectrum
\begin{align}\label{eq:SP}
\begin{split}
\rho(s) =\lambda \delta(s-M_H^2)+ \rho_{\rm{cont}}(s),\
\end{split}
\end{align}
where $\rho_{\rm{cont}}(s)$ denotes the continuum spectral density.

However, in the region where $-q^2=Q^2\gg\Lambda^2_{QCD}$, correlation function $\Pi(q^2)$ can be calculated by using the OPE, which reads
\begin{align}\label{eq:OPE}
\begin{split}
\Pi(q^2) &=C_1(q^2) +\sum_i C_i(q^2) \langle O_i \rangle \, ,
\end{split}
\end{align}
where $C_{1}$ and $C_i$ are the perturbatively calculable Wilson coefficients for perturbative term $\langle \Omega|1 |\Omega \rangle=1 $ and vacuum condensation $\langle O_i  \rangle=\langle \Omega|O_i |\Omega \rangle $, respectively. The relative importance of the vacuum condensation is power suppressed by the dimension of operator $O_i$. In our calculations, we only maintain the vacuum condensations up to dimension d=4, which gives the approximation expression of the OPE as
\begin{align}\label{eq:OPE2}
\begin{split}
\Pi(q^2) & = C_1(q^2)+ C_{GG}(q^2) \langle g_s^2 \hat{G}\hat{G} \rangle \, ,
\end{split}
\end{align}
where $\langle g_s^2 \hat{G}\hat{G} \rangle$ denotes the gluon-gluon (GG) condensation. Here, the contributions of higher dimensional condensates are expected to be small because of the power suppressions. Particularly, the d=6 term $\langle g_s^3 \hat{G}\hat{G}\hat{G} \rangle$ is neglected, and this is illustrated by a recent work\cite{Wang:2020avt}, where the contribution of this operator is shown to be negligible.

According to Eq.~(\ref{eq:K-LSR}), one can relate the physical spectrum density to the imaginary part of $\Pi(q^2)$ in Eq.~(\ref{eq:OPE2}) using the dispersion relation, which gives
\begin{align}\label{eq:KL-DisInt}
\begin{split}
\Pi(q^2) & =  \int {ds \frac{\rho(s)}{s-q^2-i\epsilon}} \, \\
&= \frac{1}{\pi} \int_{s_{\rm{th}}}^\infty ds \frac{{\rm{Im}} C_1(s)+ {\rm{Im}} C_{GG}(s)\langle g_s^2 \hat{G}\hat{G} \rangle }{s-q^2-i\epsilon} \,,
\end{split}
\end{align}
where $s_{\rm{th}}=9 m_Q^2$ is the QCD threshold for the $QQQ$ system, and the integral in the second line is assumed to be convergent.

To extract the mass of the ground state, we first, employ the quark-hadron duality \cite{Shifman:2000jv,Colangelo:2000dp,Shifman:2001qm,Narison:2007spa} and the Borel transformation, and obtain a sum rule for  $\Pi(q^2)$,
\begin{align}\label{eq:Borel-QHD}
\begin{split}
\lambda \  e^{-\frac{M_H^2}{M_B^2}} =&  \int_{s_{\rm{th}}}^{s_0} ds \frac{1}{\pi}{\rm{Im}} C_1(s)\, e^{-\frac{s}{M_B^2}}  \,+\int_{s_{\rm{th}}}^\infty ds \frac{1}{\pi}{\rm{Im}} C_{GG}(s) e^{-\frac{s}{M_B^2}}\langle g_s^2 \hat{G}\hat{G} \rangle\,,
\end{split}
\end{align}
where $s_0$ and $M_B$ are the continuum threshold and Borel parameters respectively, which are introduced here owing to the qurak-hadron duality and Borel transformation. Differentiating both sides of Eq.~(\ref{eq:Borel-QHD}) with respect to $-\frac{1}{M_B^2}$, we obtain
\begin{align}\label{eq:diff}
\begin{split}
\lambda\  M_H^2\  e^{-\frac{M_H^2}{M_B^2}} =&  \int_{s_{\rm{th}}}^{s_0} ds \frac{1}{\pi}{\rm{Im}} C_1(s)\, e^{-\frac{s}{M_B^2}} s +\int_{s_{\rm{th}}}^\infty ds \frac{1}{\pi}{\rm{Im}} C_{GG}(s) e^{-\frac{s}{M_B^2}} s\langle g_s^2 \hat{G}\hat{G} \rangle\,.
\end{split}
\end{align}
Finally, we can solve $M_H$ using Eq.~(\ref{eq:Borel-QHD}) and (\ref{eq:diff}),
\begin{align}\label{eq:MH}
\begin{split}
M_H^2 &= \frac{\int_{s_{\rm{th}}}^{s_0} ds\, \, \rho_{1}(s)\, e^{-\frac{s}{M_B^2}} s+\int_{s_{\rm{th}}}^\infty ds \rho_{GG}(s) e^{-\frac{s}{M_B^2}} s\langle g_s^2 \hat{G}\hat{G} \rangle}{\int_{s_{\rm{th}}}^{s_0} ds\, \, \rho_{1}(s)\, e^{-\frac{s}{M_B^2}}+\int_{s_{\rm{th}}}^\infty ds \rho_{GG}(s) e^{-\frac{s}{M_B^2}}\langle g_s^2 \hat{G}\hat{G} \rangle}\,,
\end{split}
\end{align}
where $\rho_1=\frac{1}{\pi}{\rm{Im}} C_1$ and $\rho_{GG}=\frac{1}{\pi}{\rm{Im}} C_{GG}$.


\section{Calculations of $C_1$ and $C_{GG}$ for $QQQ$ baryon}\label{sec:Calculation of C1 and CGG}

In QCD sum rules, there are two kinds of expansions, the OPE and perturbative expansion in $\alpha_s$. For the OPE, we consider the most important contributions, i.e., perturbative term $C_1$ and the GG condensation $C_{GG} \langle g_s^2 \hat{G}\hat{G} \rangle$, because other higher dimensional operators are power suppressed. According to Eq.~(\ref{eq:MH}), we need the imaginary parts of $C_1$ and $C_{GG}$, which can be calculated perturbatively.

\begin{figure}[H]
\centering
\subfigure[$C_1$-LO]{
\includegraphics[scale=1]{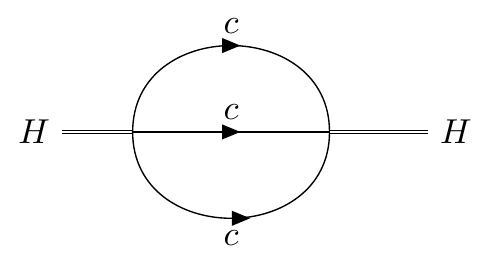}
}
\subfigure[$C_{GG}$-LO]{
\includegraphics[scale=1]{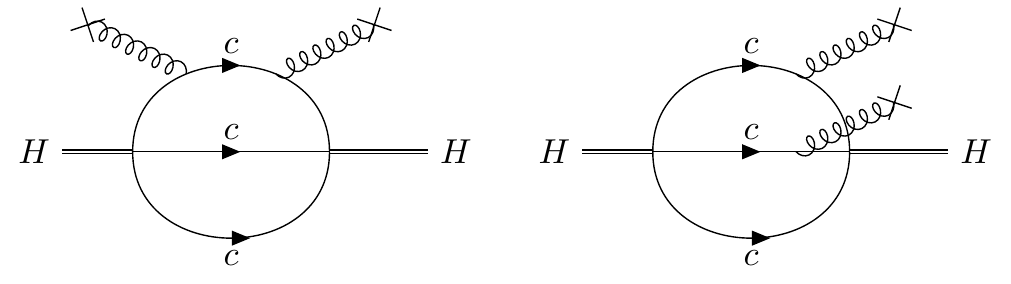}
}
\caption{\label{fig:FeynmanDiagrams-LO}
  LO Feynman Diagrams of $C_1$ and $C_{GG}$. $H$ denotes the interpolating current.}
\end{figure}

\begin{figure}[H]
\centering
\subfigure[$C_1$-NLO]{
\includegraphics[scale=1]{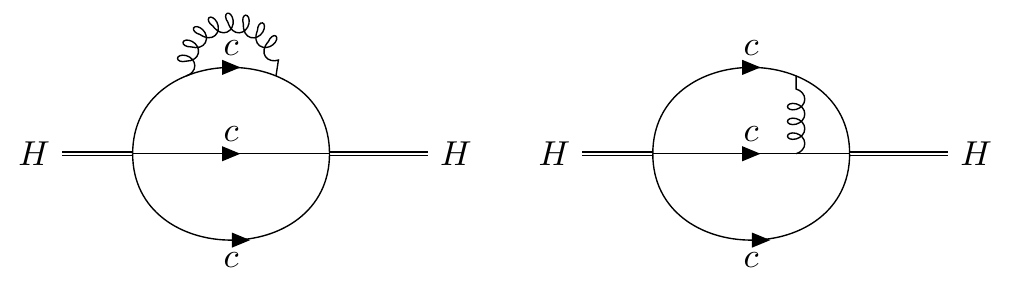}
}
\subfigure[$C_1$-NLOct]{
\includegraphics[scale=1]{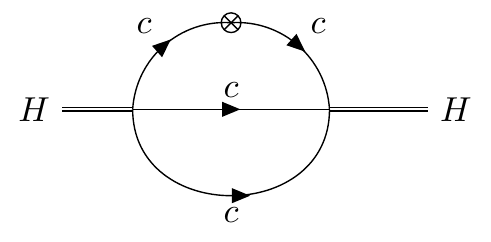}
}
\caption{\label{fig:FeynmanDiagrams-NLO}
  NLO and counter term Feynman Diagrams of $C_1$. $H$ denotes the interpolating current.}
\end{figure}

We use FeynArts \cite{Kublbeck:1990xc,Hahn:2000kx} to generate the Feynman diagrams and amplitudes of $C_1$ and $C_{GG}$. The LO and NLO Feynman diagrams are shown in Fig.~\ref{fig:FeynmanDiagrams-LO} and Fig.~\ref{fig:FeynmanDiagrams-NLO}, respectively.

In the $\Omega_{QQQ}$ system, the three quarks have the same flavor, thus some diagrams are similar and not shown again. For example, there are three cases in the second diagram of $C_1$-NLO in Fig.\ref{fig:FeynmanDiagrams-NLO}, which denotes the cases with one gluon exchange between any two heavy quarks.

The calculation procedures for $C_1$ and $C_{GG}$ are summarized below:
\begin{itemize}
\item
  1. We use FeynCalc \cite{Mertig:1990an,Shtabovenko:2016sxi} to simplify the spinor structures of the Feynman amplitudes with the Naive-$\gamma_5$ scheme.
\item
  2. We use Reduze \cite{vonManteuffel:2012np} to reduce all the loop integrals $\mathbf{I}$ to the linear combination $\mathbf{I} =\sum a_i\,  \tilde{I}_i$, where $\tilde{I}_i$'s are the so-called master integrals (MIs). Furthermore, $a_i$s are the reduced coefficients, which are rational functions in the Mandelstam invariant (s), quark mass ($m_Q$) and time-space dimension ($D=4-2\epsilon$) in this work.
 \item
  3. We set up differential equations  for MIs \cite{Kotikov:1990kg,Bern:1992em,Remiddi:1997ny,Gehrmann:1999as} and solve them numerically, with boundary conditions obtained via auxiliary mass flow \cite{Liu:2017jxz}.(The expressions of the MIs are the series-expansion forms, i.e. $\tilde{I}(s,r,\epsilon)=s^{a+b\epsilon} \sum_{n,m} (\sum_{i} c_{mni} \epsilon^i)Log(r)^{m} r^{n} $, where $r=m_Q^2/s$ and $c_{mni}$s are float numbers rather than rational numbers. The expressions of the Wilson coefficients $C_1$ and $C_{GG}$ are complex and not presented in this paper. These results will be shared in auxiliary files.)
  \item
  4. Renormalization. There are no infrared divergences in the NLO amplitude of $C_1$. After performing the wave-function and mass renormalization of the quarks ($m_Q$ is renormalized in either the $\overline{\rm{MS}}$ or On-Shell scheme), the remaining ultraviolet divergences can be removed by the operator renormalization of current $J_\mu$. We renormalize the current in the $\overline{\rm{MS}}$ scheme, and find the renormalizaion constant up to the NLO level as
  \begin{equation}\label{eq:R+-meson-renormalization}
  Z_{O}=1+\frac{\alpha_s}{6 \pi }\left(\frac{1}{\epsilon}+\mbox{Log}(4\pi)-\gamma_E \right)
\end{equation}
\end{itemize}


\section{Phenomenology}\label{sec:Phenomenology}

In our numerical analysis, we choose the following parameters \cite{Bagan:1992za,Dominguez:1994ce,Dominguez:2014pga,Aoki:2016frl,Wang:2017qvg,Zyla:2020zbs}.
\begin{align}\label{eq:parameters}
\begin{split}
m_c^{\overline{\rm{MS}}}(m_c)&=1.28 \pm 0.03\, \,  {\rm{GeV}}\,\\
m_c^{OS}&=1.46 \pm 0.07\, \,  {\rm{GeV}}\,\\
m_b^{\overline{\rm{MS}}}(m_b)&=4.18 \pm 0.03\, \,  {\rm{GeV}}\,\\
m_b^{OS}&=4.65 \pm 0.05\, \,  {\rm{GeV}}\,\\
\langle g_s^2 \hat{G}\hat{G} \rangle &= 4\pi^2(0.037\pm0.015) \, \, {\rm{GeV}}^4 \,\\
\alpha_s(m_Z&=91.1876 ~{\rm{GeV}})=0.1181
\end{split}
\end{align}
It is worth emphasizing that $\alpha_s (\mu)$ and the heavy quark mass $m_Q^{\overline{\rm{MS}}}(\mu)$ are obtained through two-loop running~\cite{Wang:2017qvg}.As a typical and tentative choice, we set the renormalization scale $\mu$ to be equal to the Borel parameter $M_B$ in our phenomenological analysis \cite{Shifman:1978bx,Bertlmann:1981he} (we will discuss the renormalization scale dependence of the results by setting $\mu$ with other values later).
Moreover, the On-Shell (OS) masses $m_c^{OS}$ and $m_b^{OS}$ are extracted from the QCD sum rules analysis of the $J/\psi$ and $\Upsilon(1S)$ spectrums, respectively, in which the mass renormalization scheme and truncation order of $\alpha_s$ are the same as ours.

According to Eq.~(\ref{eq:MH}), the numerical result of $M_H$ depends on two parameters: $s_0$ and $M_B$. Principally, $M_H$ as a physical mass, is independent of any artificial parameters. Therefore a credible result should be obtained in an appropriate region of these two parameters, where $M_H$ is dependent weakly on $M_B$ and $s_0$. Additionally, the choice of $M_B$ and $s_0$ should ensure the validity of the OPE and dominance of the ground-state pole contribution, which will constrain the two parameters within a suitable parameter space, the so-called "Borel window". Within the Borel window, we find the so-called "Borel platform", in which $M_H$ weakly depends on $s_0$ and $M_B$.

To search for the Borel window, we define the relative contributions of the condensation and continuum as
\begin{align}\label{eq:borelwindow}
\begin{split}
r_{GG} &=\frac{\langle g_s^2 \hat{G}\hat{G} \rangle\int_{s_{\rm{th}}}^{\infty} ds\, \, \rho_{GG}(s)\, e^{-\frac{s}{M_B^2}}}{\int_{s_{\rm{th}}}^{\infty} ds\, \, \rho_{1}(s)\, e^{-\frac{s}{M_B^2}}}\\
r_{\rm{cont}} &=\frac{\int_{s_0}^{\infty} ds\, \, \rho_{1}(s)\, e^{-\frac{s}{M_B^2}}}{\int_{s_{\rm{th}}}^{\infty} ds\, \, \rho_{1}(s)\, e^{-\frac{s}{M_B^2}}}
\end{split}
\end{align}
and impose the following constraints:
\begin{equation}\label{eq:BWcondition-MSbar}
|r_{GG}|\leq 30\%, \qquad  |r_{\rm{cont}}|\leq 30\%
\end{equation}
The two constraints guarantee the validity of OPE and the gound-state contribution dominance, separately.
To find the Borel platform, we first search for the point on which the parameter dependence of $M_H$ is weakest within the Borel window. In addition to the conditions given in (\ref{eq:BWcondition-MSbar}), we also impose the following constrain on $s_0$:
\begin{equation}\label{eq:S0condition}
s_0<(M_H+1~\rm{GeV})^2,
\end{equation}
since $s_0$ roughly denotes the energy scale where the continuum spectrum begins and the energy level spacing of a heavy hadron system is usually smaller than 1 GeV.
More explicitly, we choose the variables as $x=s_0$ and $y=M_B^2$ and define the function describing the flatness degree as
\begin{align}\label{eq:findPlatform}
\Delta(x,y)=\left( \frac{\partial M_H}{\partial x}\right)^2+ \left ( \frac{\partial M_H}{\partial y}\right)^2\,.
\end{align}
Thus, minimizing the function $\Delta(x,y)$ within the Borel window and with the constrain (\ref{eq:S0condition}), we obtain the point ($x_0,y_0$), which will be used to evaluate the central value of $M_H$. We then vary the values of $s_0$ and $M_B^2$ around the point ($x_0,y_0$) up to a 10\% magnitude to estimate the errors of $M_H$. It should be emphasized that the central point ($x_0,y_0$) may lie on the margin of the Borel window in some cases, therefore, the parameter space used to estimate the errors of $M_H$ may exceed the Borel window. Additionally, the upper and lower errors are usually not symmetric.

\subsection{$\Omega_{ccc}$ System}\label{subsec:Omega_ccc}

\begin{table}[H]
  \renewcommand\arraystretch{2}
  \begin{center}
  \setlength{\tabcolsep}{2 mm}
\begin{tabular}{ccccccccc}
 \hline\hline
  Order&   $M_H$ (GeV)   &  $s_0$ (${\rm{GeV}}^2$)  & $M_B^2$ (${\rm{GeV}}^2$)  & \makecell{Error from \\$s_0$ and $M_B^2$ } & \makecell{Error from \\$m_Q$ } & \makecell{Error from \\$\mu$}\\ \hline

LO($\overline{\rm{MS}}$) &    $4.39^{+0.24}_{-0.22}$   &    $29(\pm 10\%)$   &   $1.75(\pm 10\%)$   &  $^{+0.02}_{-0.03}$ &  $^{+0.10}_{-0.10}$ & $^{+0.22}_{-0.19}$ \\

NLO($\overline{\rm{MS}}$) &   $4.53^{+0.26}_{-0.11}$   &    $26(\pm 10\%)$   &   $1.20(\pm 10\%)$    &  $^{+0.02}_{-0.04}$ &  $^{+0.09}_{-0.10}$ &  $^{+0.24}_{-0.00}$\\

LO(OS)     &         $4.51^{+0.19}_{-0.23}$         &      $21(\pm 10\%)$    &   $0.40(\pm 10\%)$    & $^{+0.06}_{-0.15}$ &  $^{+0.18}_{-0.18}$ \\

NLO(OS)     &         $4.55^{+0.13}_{-0.18}$           &   $22(\pm 10\%)$     &   $0.90(\pm 10\%)$    & $^{+0.07}_{-0.12}$ &  $^{+0.11}_{-0.14}$ \\
\hline\hline
   \end{tabular}

   \caption{LO and NLO results for the mass of $\Omega_{ccc}^{++}$ in $\overline{\rm{MS}}$ and On-Shell schemes. Here, the errors for $M_H$ are from $s_0, M_B$, the charm quark mass, and the renormalization scale $\mu$ with $\mu=kM_B$ and $k \in (0.8, 1.2)$ (the central values correspond to $\mu=M_B$ ).}
   \label{tab:R-NLOresult-MSbar-OS}
   \end{center}
   \end{table}

\begin{figure}[H]
\centering
\subfigure[$\overline{\rm{MS}}$]{
\includegraphics[scale=0.45]{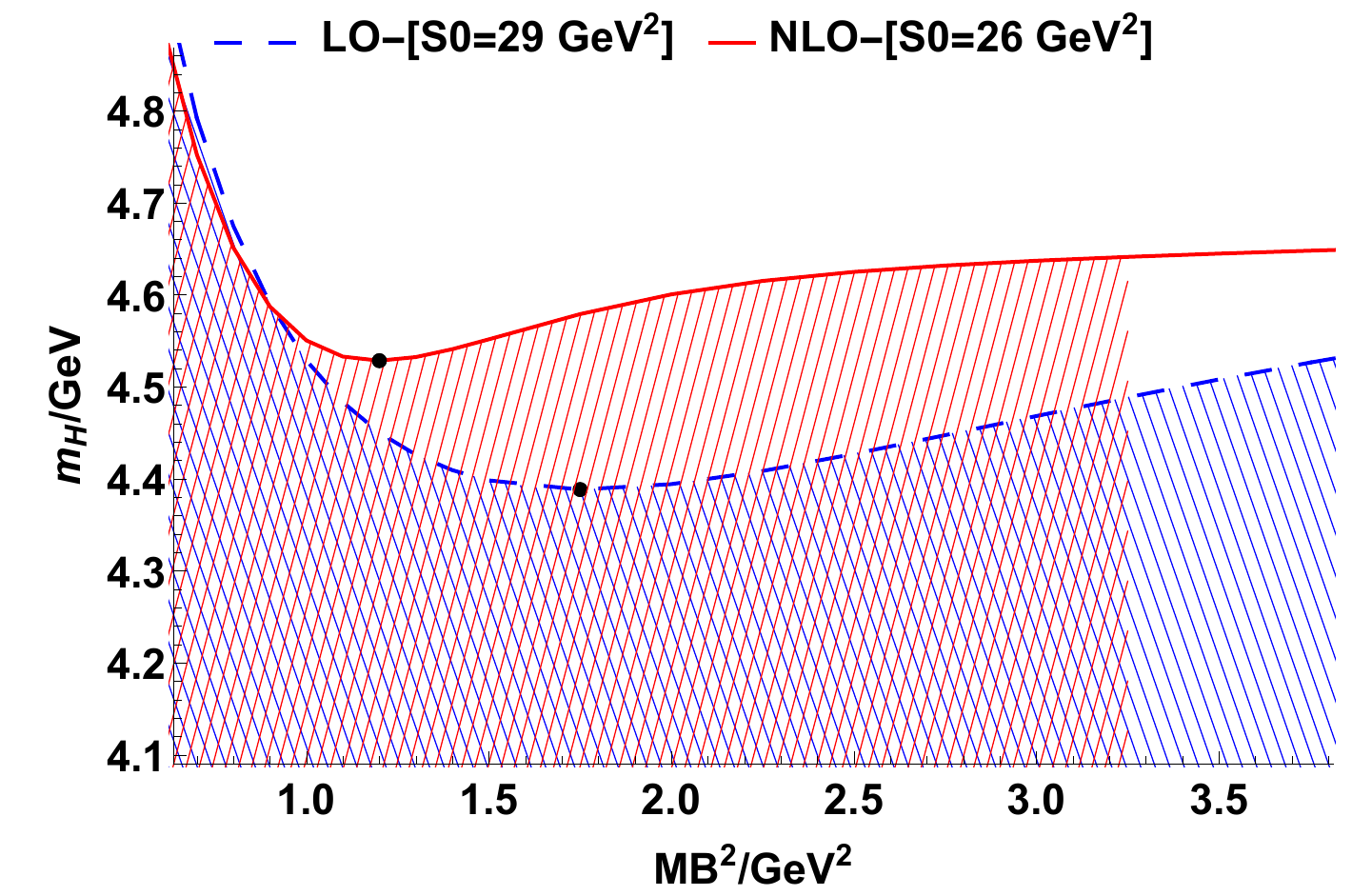}
\includegraphics[scale=0.45]{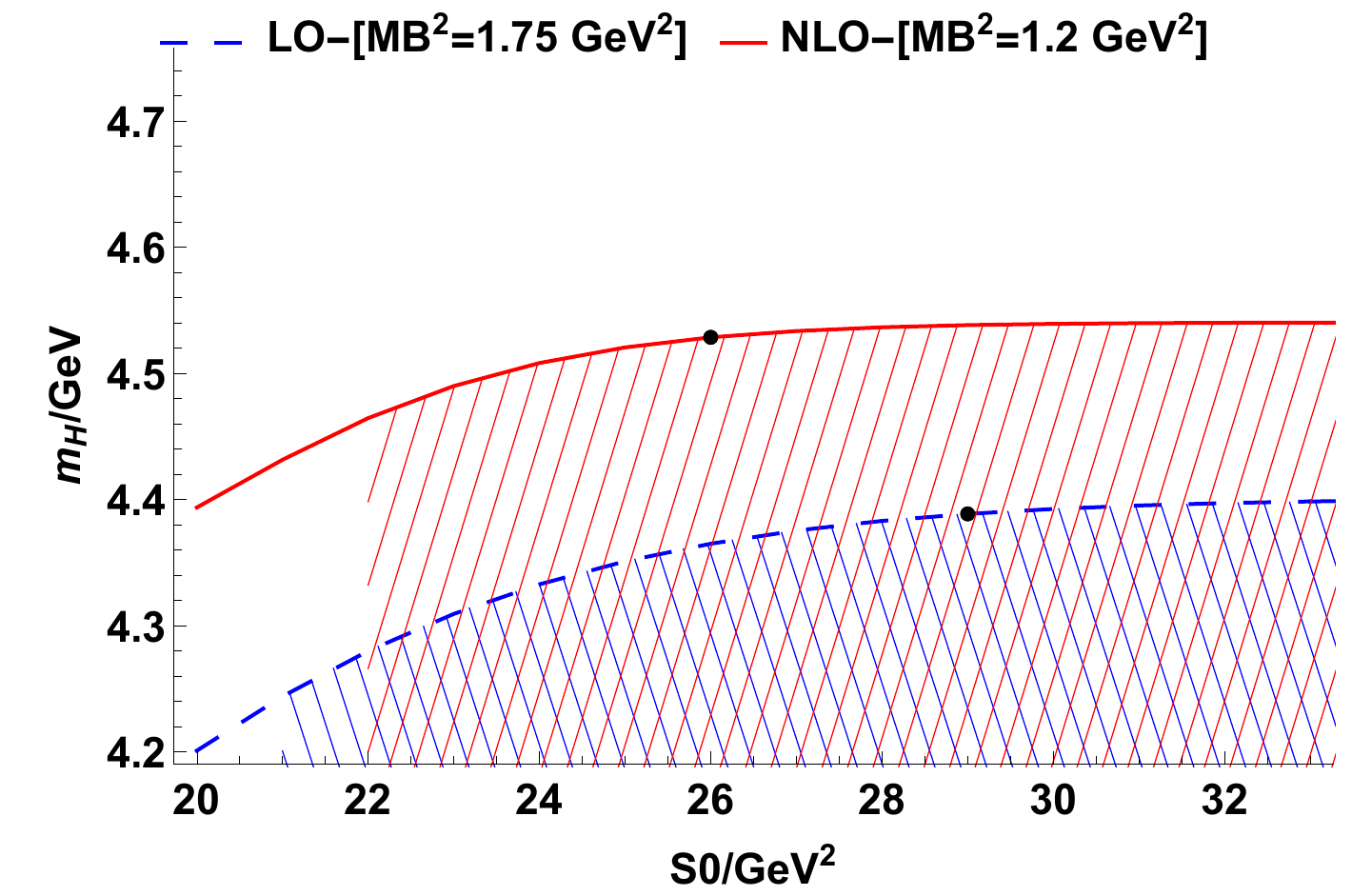}
}\\
\subfigure[OS]{
\includegraphics[scale=0.45]{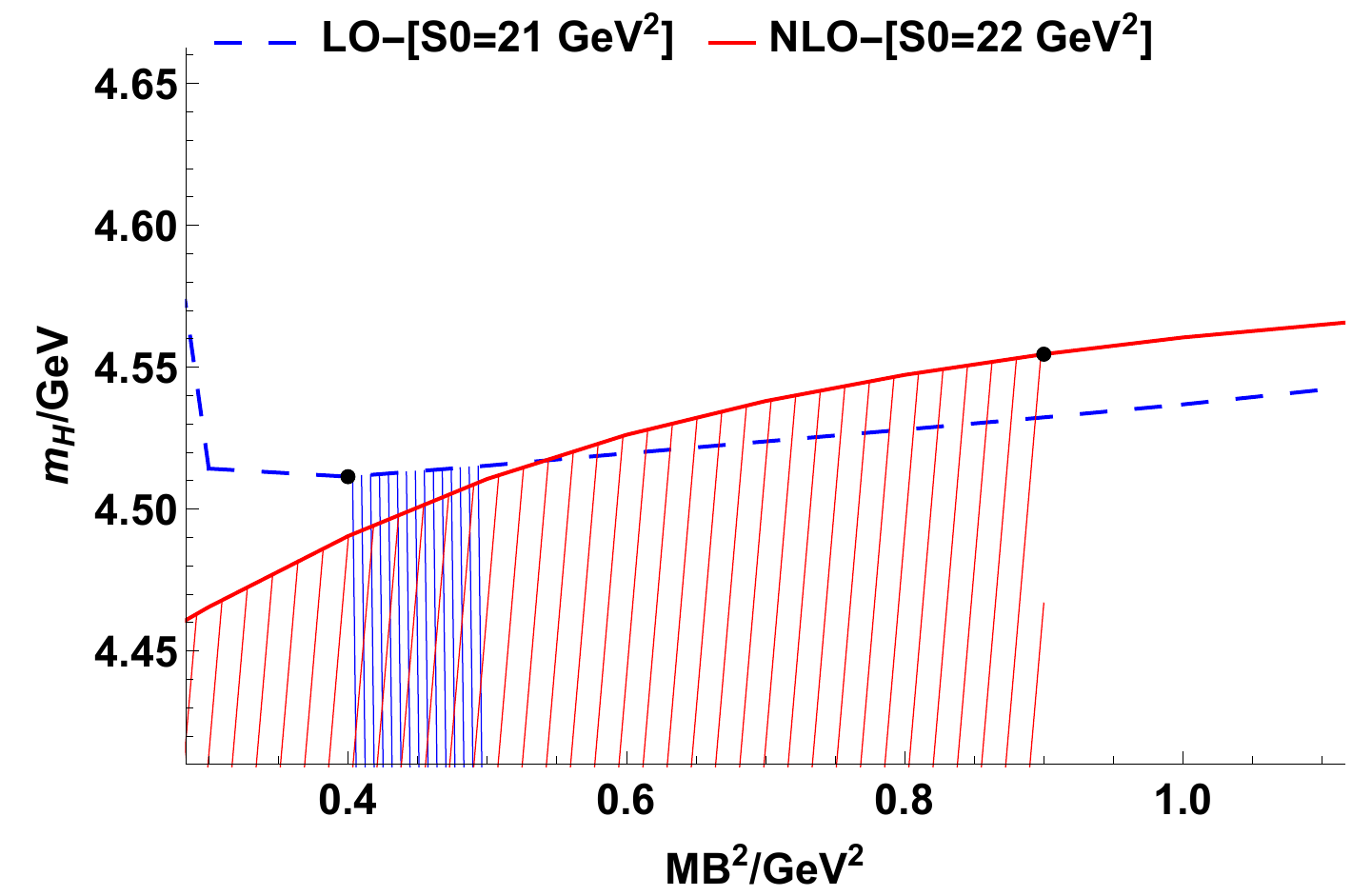}
\includegraphics[scale=0.45]{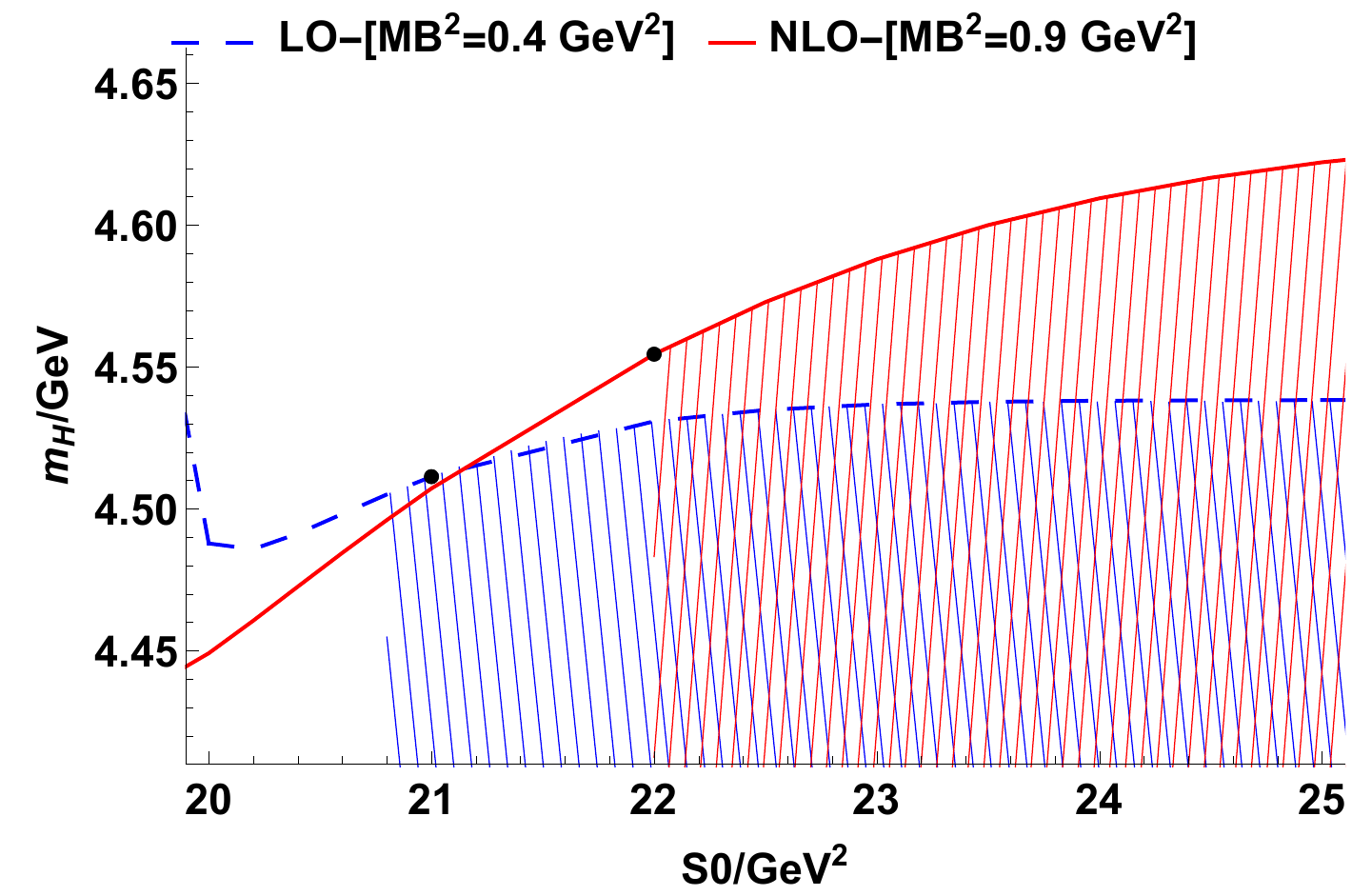}
}
\caption{\label{fig:R-NLO-MSbar-OS}
The Borel platform curves for the $\Omega_{ccc}^{++}$ in $\overline{\rm{MS}}$ and On-Shell schemes.}
\end{figure}

The results for $\Omega_{ccc}^{++}$ in the $\overline{\rm{MS}}$ and OS schemes are listed in Table~\ref{tab:R-NLOresult-MSbar-OS}. The Borel platform curves, which show the parameter dependence of $M_H$ on $s_0$ and $M_B^2$, are shown in Fig.~\ref{fig:R-NLO-MSbar-OS}. In all the curves, the dot corresponds to the central point ($x_0,y_0$), and the shadows denote the Borel windows determined by Eq.~(\ref{eq:BWcondition-MSbar})

First, Fig.~\ref{fig:R-NLO-MSbar-OS}, we observe that there exists a perfect Borel platform in the $\overline{\rm{MS}}$ scheme. However there is no decent Borel platform in the OS scheme, thus the result in the OS scheme is not good. Therefore, we consider the result in the $\overline{\rm{MS}}$ scheme as our prediction for the mass of $\Omega_{ccc}^{++}$.

Secondly, although the OS result is not good, we can still observe the distinct NLO effects in the reduction of the scheme dependence by the comparison between the $\overline{\rm{MS}}$ and OS schemes. In fact, at the LO a big gap is observed between the two schemes results, whereas the $\overline{\rm{MS}}$ result increases markedly at the NLO, which brings the results of the two schemes close to each other at the NLO. Furthermore, the quark mass dependence of the results is also reduced at the NLO. Particularly, the NLO contribution leads to a noteworthy and significant correction of the LO result, and cannot be ignored in the QCD sum rules for the triply charmed baryons.

\begin{figure}[H]
\centering
\includegraphics[scale=0.4]{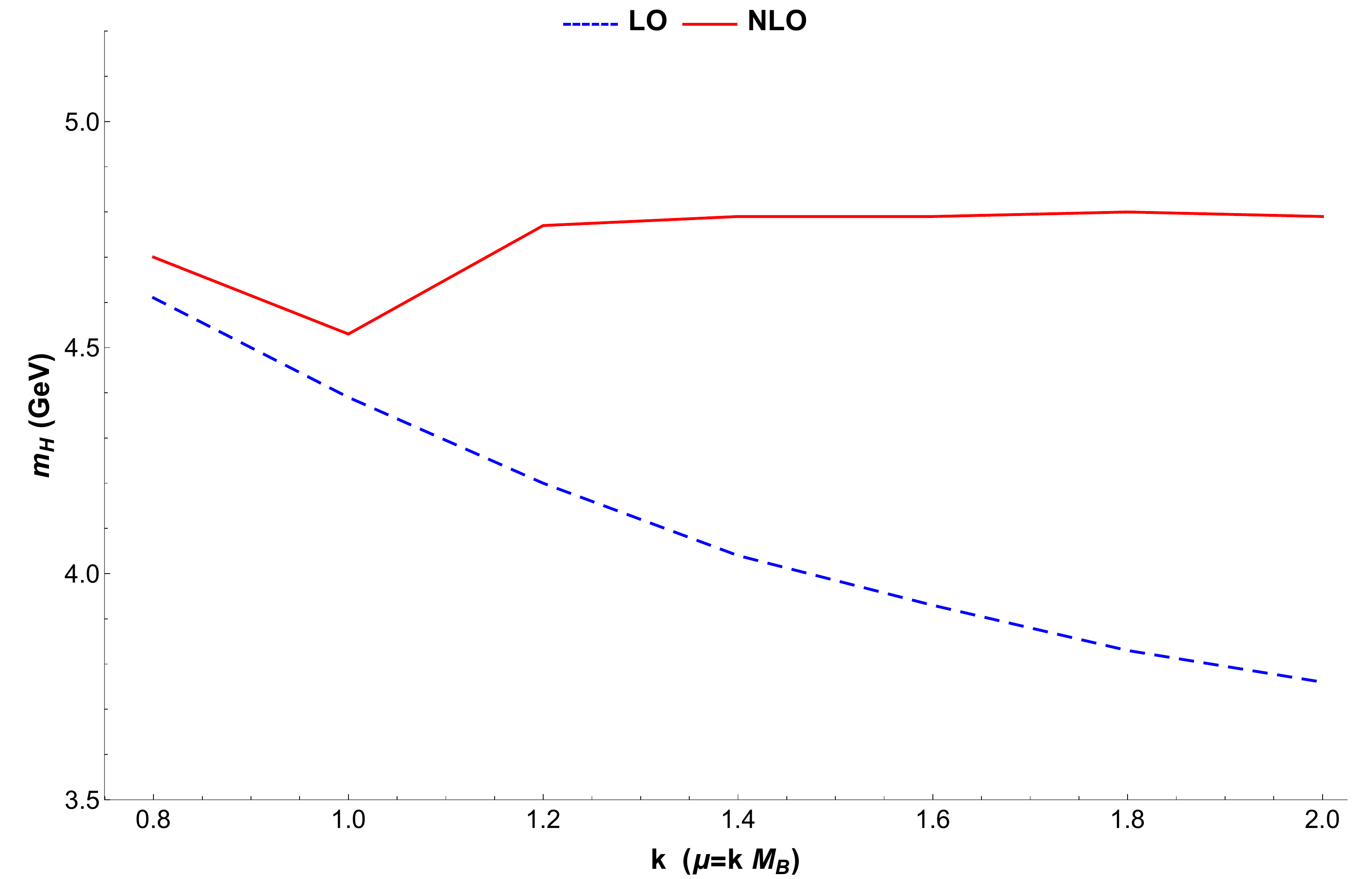}
\caption{\label{fig:3c-mu--kMB}
 Dependence on renormalization scale $\mu$ in $\overline{\rm{MS}}$ scheme for $\Omega_{ccc}^{++}$.}
\end{figure}

Furthermore, as already mentioned, we will evaluate the renormalization scale  $\mu$ dependence of the LO and NLO results in the $\overline{\rm{MS}}$ scheme. The results obtained with $\mu=M_B$ are listed in Table I. We then study the results with different $\mu$ ($\mu=k \ M_B$) and $k \in (0.8, 2.0)$, which are presented in Fig.[\ref{fig:3c-mu--kMB}] and  Tab.~\ref{tab:3c-mudependence-data}. The range of $\mu$ is chosen with the requirement that the Borel platform can be achieved and the perturbative expansion is under good control. From Fig.[\ref{fig:3c-mu--kMB}] it can be clearly observed that the $\mu$ dependence is significantly reduced for the NLO result, whereas the LO result is very sensitive to the choice of $\mu$. In addition, it is very interesting to note that the obtained NLO mass of $\Omega_{ccc}^{++}$ is 4.75-4.80 GeV for a rather wide range of  $\mu=(1.2-2.0) M_B$. Considering all important factors, like $\mu$ dependence and perturbative convergence, 4.75-4.80 GeV is a more credible prediction for the $\Omega_{ccc}$ mass. Interestingly enough, this value is in good agreement with the lattice QCD result.
\begin{table}
  \renewcommand\arraystretch{1.5}
  \begin{center}
  \setlength{\tabcolsep}{6 mm}
   \caption{$\overline{\rm{MS}}$ mass of $\Omega_{ccc}$ with different renormalization scale $\mu=k M_B$ }
\begin{tabular}{lc|@{\ }|c}
 \hline\hline
    k ($\mu=k \ M_B$ )   &    $LO$ (GeV) & $NLO$ (GeV)\\ \hline

0.8 &     4.65     &   4.70 \\
1.0 &     4.39    &   4.53  \\
1.2 &     4.20    &   4.77   \\
1.4 &     4.05     &        4.79  \\
1.6 &     3.93     &         4.79  \\
1.8 &    3.83    &         4.80  \\
2.0 &     3.76   &   4.79   \\
\hline\hline

   \end{tabular}
   \label{tab:3c-mudependence-data}
   \end{center}
   \end{table}

It is worth mentioning that our result includes errors due to the variation of the renormalization scale, as compared with previous LO results \cite{Zhang:2009re,Wang:2020avt} ($4.67 \pm0.15~{\rm{GeV}}$ and $4.81 \pm0.10~{\rm{GeV}}$, respectively, in which the scale dependence is not considered).
Regarding the central values, the differences between our LO results and those in \cite{Zhang:2009re,Wang:2020avt} are mainly caused by the choice of input parameters and determination Borel platform. For example, we can reproduce the result in \cite{Wang:2020avt} by using the parameters therein. In Tab.~\ref{tab:3c-3b-DiffModels}, we list the predictions for the mass of $\Omega_{ccc}^{++}$ in various approaches.

\subsection{$\Omega_{bbb}$ System}\label{subsec:Omega_bbb}

\begin{figure}[H]
\centering
\subfigure[$\overline{\rm{MS}}$]{
\includegraphics[scale=0.45]{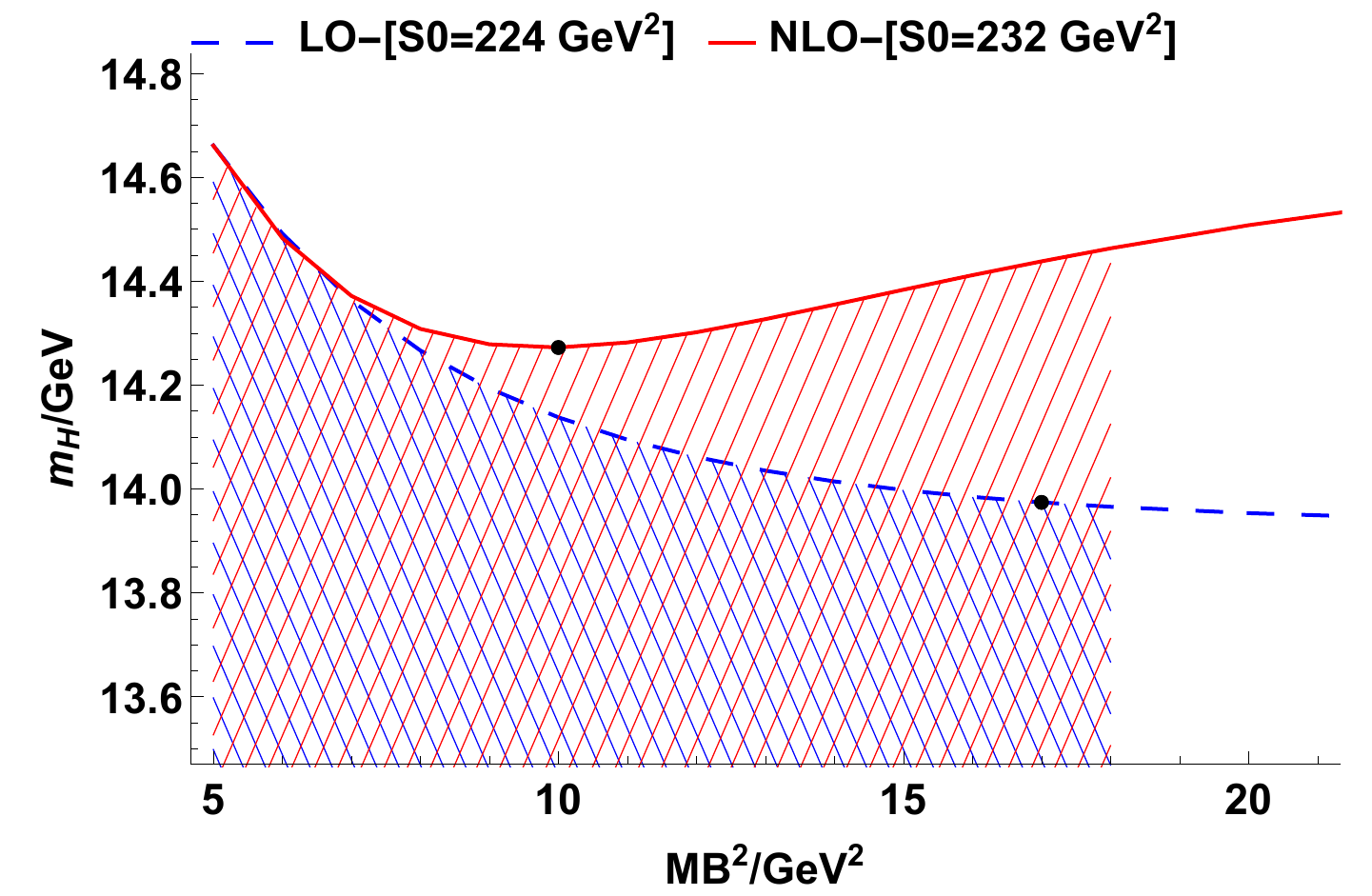}
\includegraphics[scale=0.45]{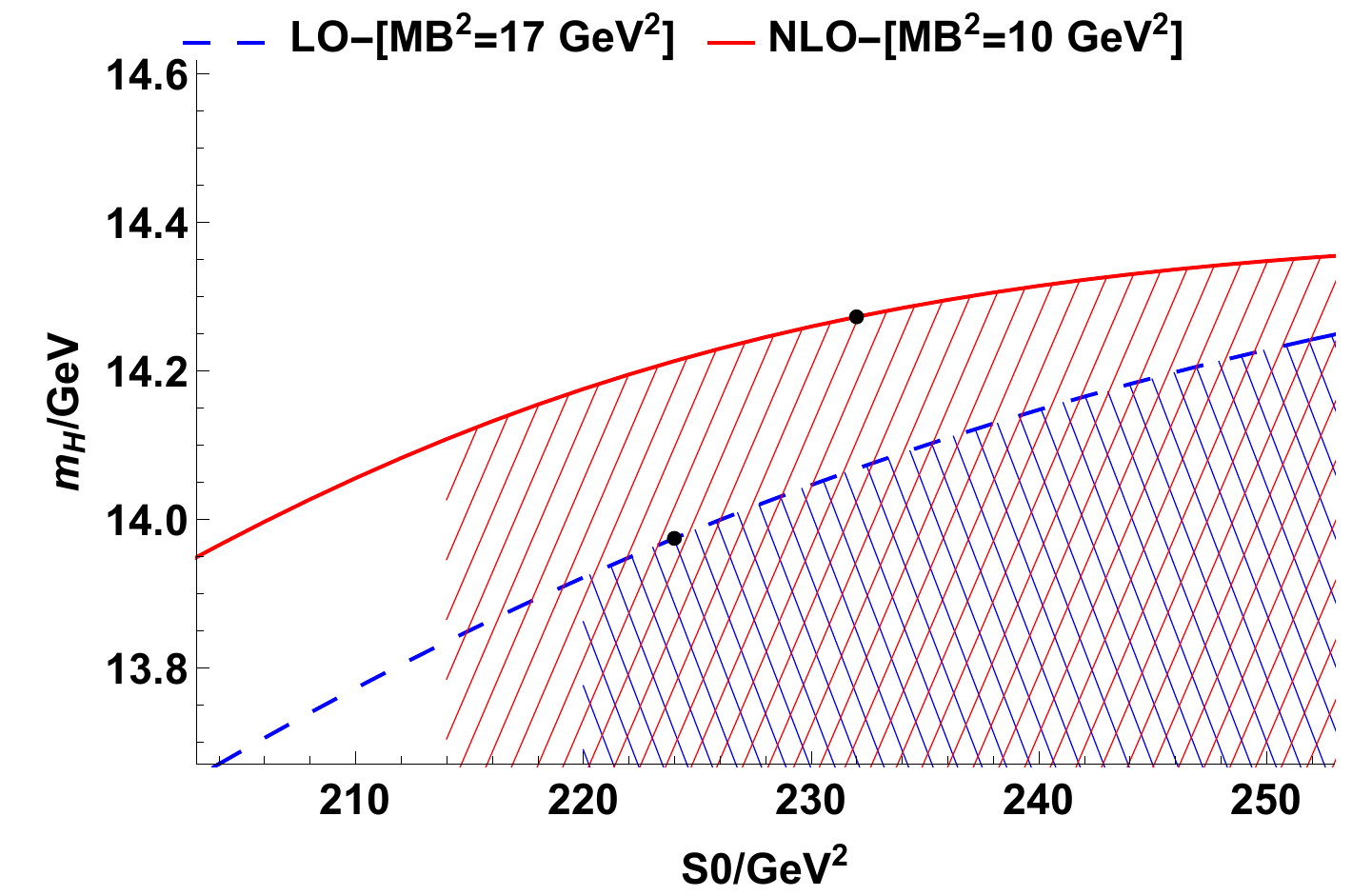}
}\\
\subfigure[OS]{
\includegraphics[scale=0.45]{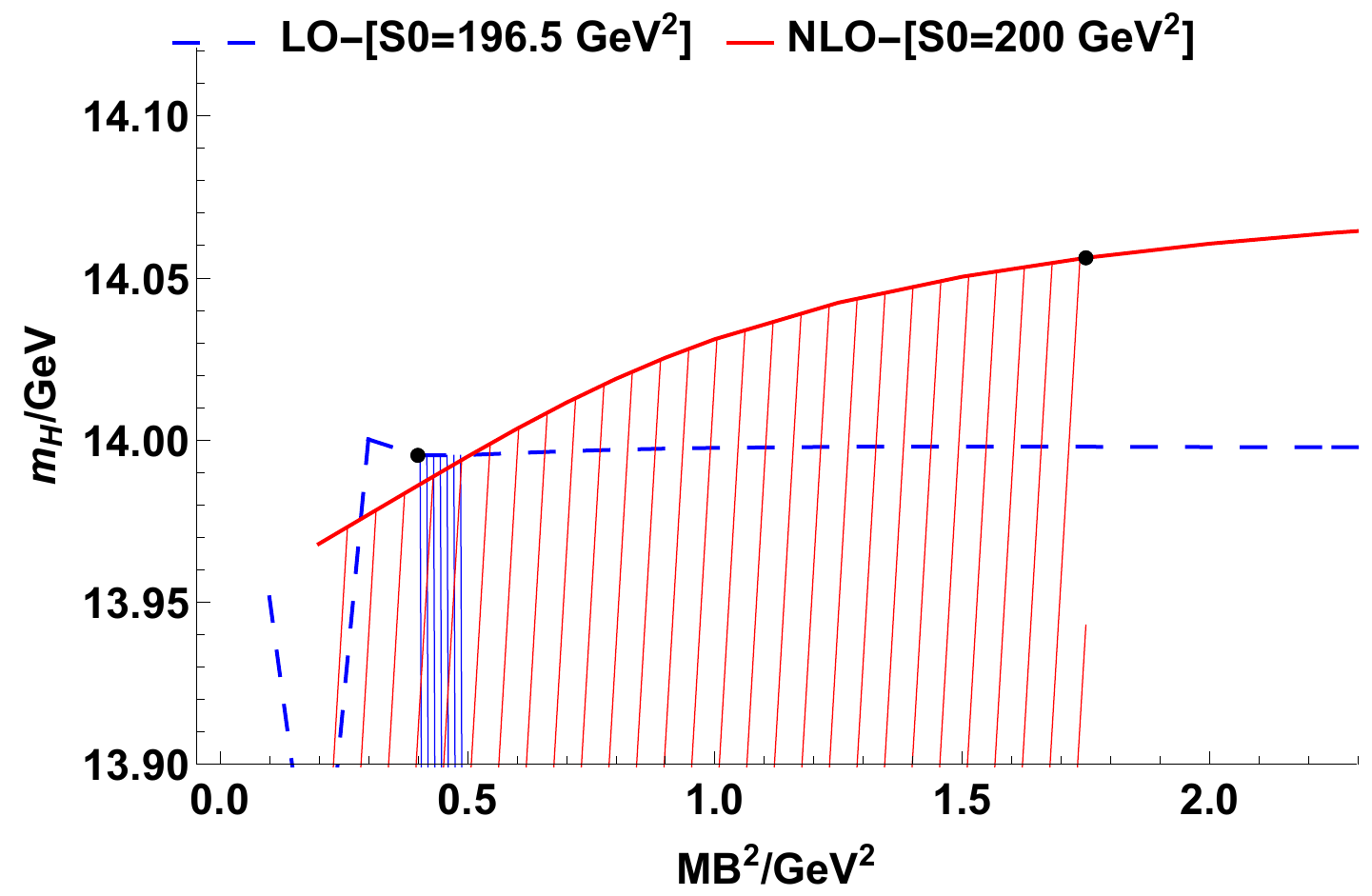}
\includegraphics[scale=0.45]{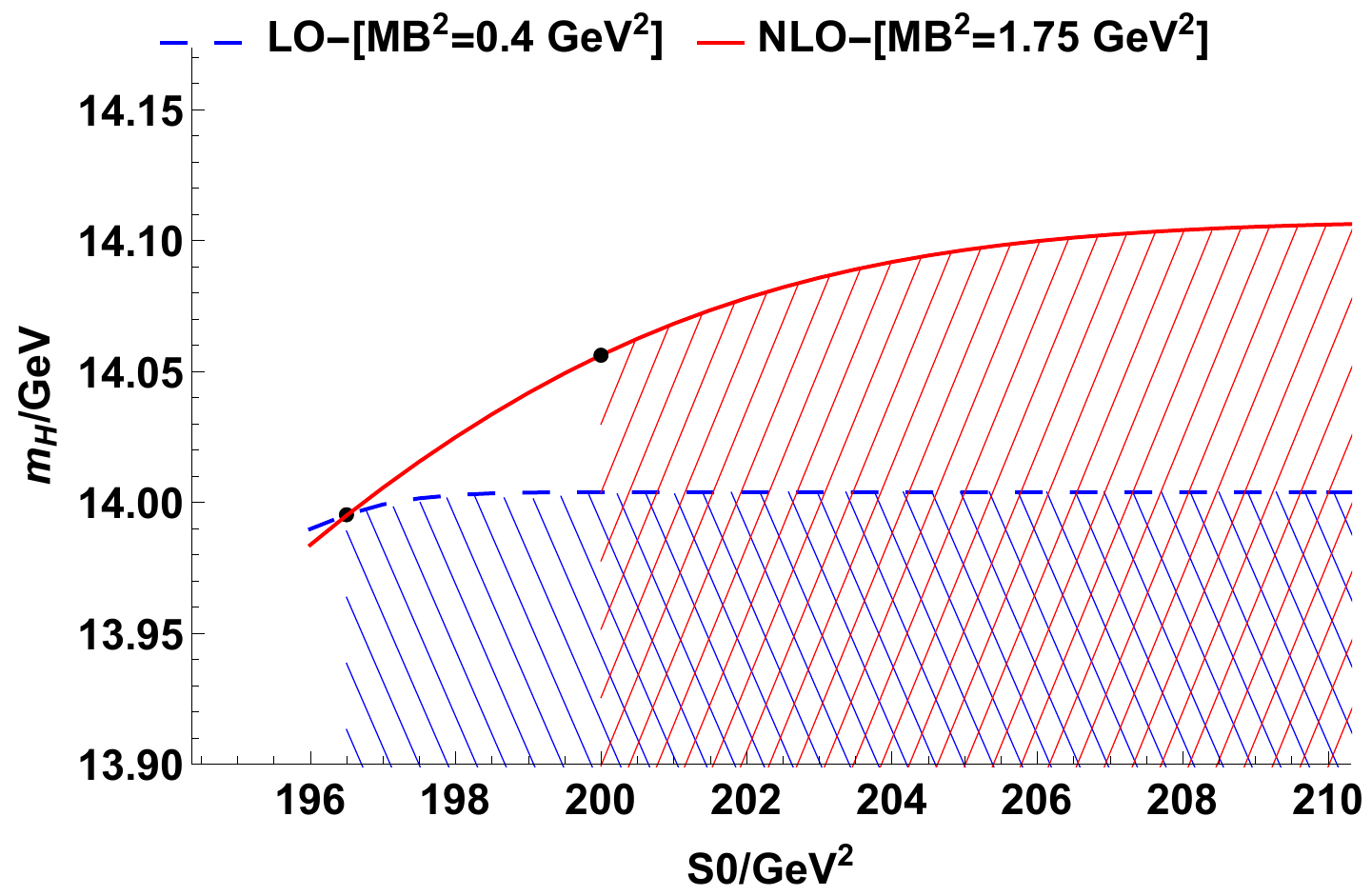}
}
\caption{\label{fig:R-NLO-MSbar-OS-3b}
Borel platform curves for the $\Omega_{bbb}^{-}$ in $\overline{\rm{MS}}$ and On-Shell schemes.}
\end{figure}

For the $\Omega_{bbb}^{-}$, the results  in the $\overline{\rm{MS}}$ and OS schemes are listed in Table~\ref{tab:R-NLOresult-MSbar-OS-3b}, and the Borel platform curves are shown in Fig.~\ref{fig:R-NLO-MSbar-OS-3b}.

\begin{table}[H]
  \renewcommand\arraystretch{2}
  \begin{center}
  \setlength{\tabcolsep}{2 mm}
\begin{tabular}{ccccccccc}
 \hline\hline
  Order&   $M_H$ (GeV)   &  $s_0$ (${\rm{GeV}}^2$)  & $M_B^2$ (${\rm{GeV}}^2$)  & \makecell{Error from \\$s_0$ and $M_B^2$ } & \makecell{Error from \\$m_Q$ } & \makecell{Error from \\$\mu$}\\ \hline

LO($\overline{\rm{MS}}$) &    $13.97^{+0.54}_{-0.50}$   &     $224(\pm 10\%)$     &   $17.00(\pm 10\%)$    &  $^{+0.23}_{-0.36}$ &  $^{+0.06}_{-0.06}$  &$^{+0.48}_{-0.34}$ \\

NLO($\overline{\rm{MS}}$) &   $14.27^{+0.33}_{-0.32}$     &     $232(\pm 10\%)$    &   $10.00(\pm 10\%)$   &  $^{+0.12}_{-0.25}$ &  $^{+0.08}_{-0.08}$& $^{+0.30}_{-0.19}$ \\

LO(OS)     &        $14.00^{+0.15}_{-0.15}$          &     $197(\pm 10\%)$     &     $0.40(\pm 10\%)$   & $^{+0.01}_{-0.01}$ &  $^{+0.15}_{-0.15}$ \\

NLO(OS)     &        $14.06^{+0.10}_{-0.14}$      &    $200(\pm 10\%)$        &     $1.75(\pm 10\%)$   & $^{+0.07}_{-0.07}$ &  $^{+0.07}_{-0.12}$ \\
\hline\hline
   \end{tabular}

   \caption{LO and NLO results for the mass of $\Omega_{bbb}^{-}$ in $\overline{\rm{MS}}$ and On-Shell schemes. Here, the errors for $M_H$ are from $s_0, M_B$, the bottom quark mass, and renormalization scale $\mu$ with $\mu=kM_B$ and $k \in (0.8, 1.2)$ (the central values correspond to $\mu=M_B$ ).}
   \label{tab:R-NLOresult-MSbar-OS-3b}
   \end{center}
   \end{table}

With these figures, we can make similar conclusions as illustrated in the $ccc$ section. For instance, there are no very stable Borel platforms in the OS scheme at both the LO and NLO. Furthermore, there is a  distinct difference between the $bbb$ and $ccc$ systems.   This can be observed in Fig.~\ref{fig:R-NLO-MSbar-OS-3b}, where there is no Borel platform at the LO in the $\overline{\rm{MS}}$ scheme. However, there appears a clear platform after including the NLO contribution. For the $ccc$, there are platforms at both LO and NLO.
This indicates that for the $bbb$ the NLO contribution from the perturbative term $C_1$ is crucial to the formation of a stable Borel platform in the QCD sum rules.

\begin{table}[H]
  \renewcommand\arraystretch{1.5}
  \begin{center}
  \setlength{\tabcolsep}{6 mm}
   \caption{$\overline{\rm{MS}}$ mass of $\Omega_{bbb}$ with different renormalization scale $\mu=k M_B$ }
\begin{tabular}{lc|@{\ }|c}
 \hline\hline
    k ($\mu=k \ M_B$ )   &    $LO$ (GeV) & $NLO$ (GeV)\\ \hline

0.8 &     14.45     &   14.57 \\
1.0 &     13.97    &   14.27  \\
1.2 &     13.63    &   14.08 \\
1.4 &     13.34     &        13.96  \\
\hline\hline

   \end{tabular}
   \label{tab:3b-mudependence-data}
   \end{center}
   \end{table}

\begin{figure}[H]
\centering
\includegraphics[scale=0.5]{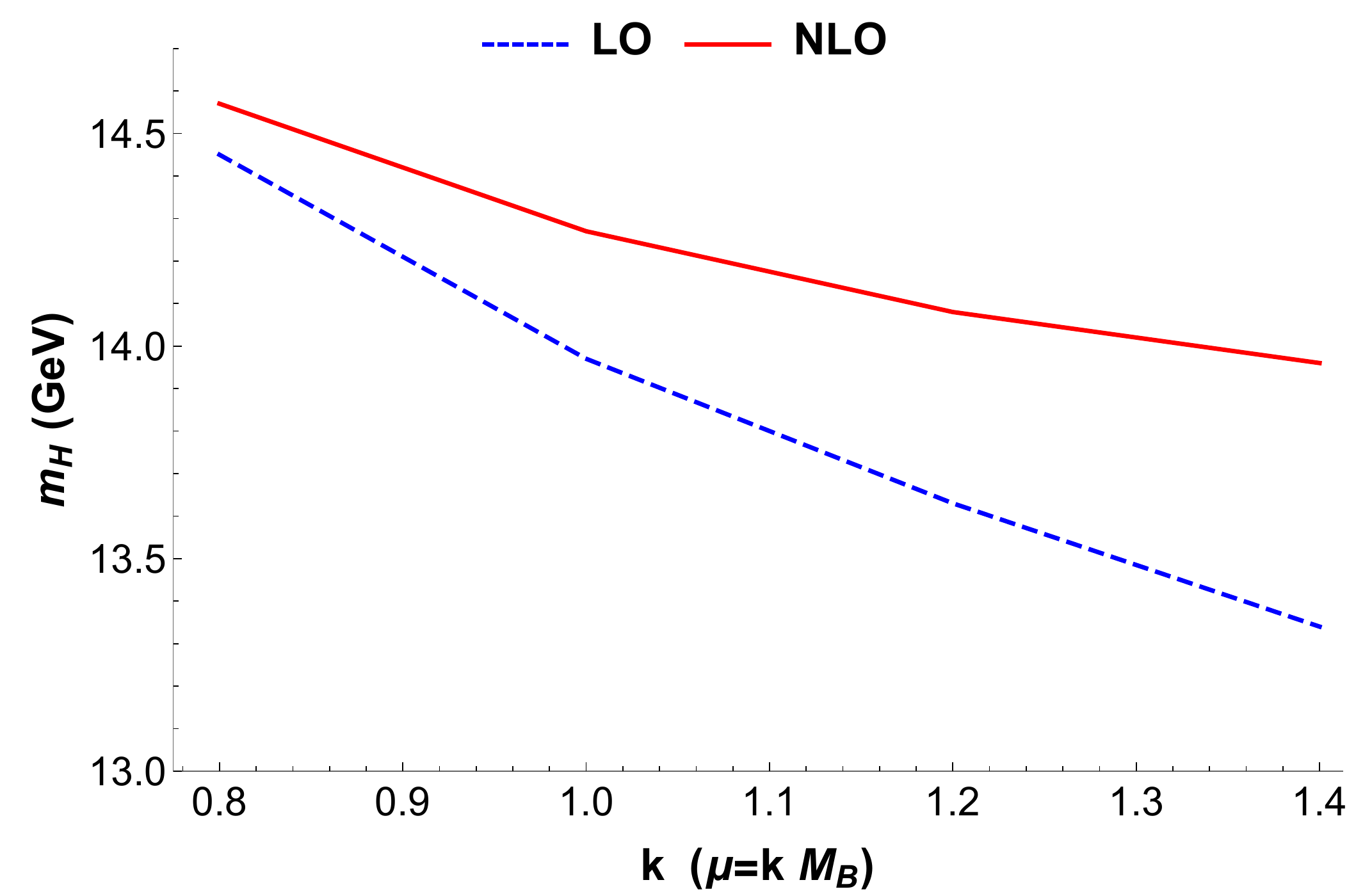}
\caption{\label{fig:3b-mu--kMB}
 The dependence on renormalization scale $\mu$ in $\overline{\rm{MS}}$ scheme for $\Omega_{bbb}^{-}$.}
\end{figure}

We further evaluate the renormalization scale $\mu$ dependence of the $\Omega_{bbb}^{-}$ mass in a wider range of $\mu=(0.8-1.4)M_B$ ($\mu>1.4 M_B$ is excluded owing to the absence of the Borel platform), and the results are shown in Fig.\ref{fig:3b-mu--kMB} and Tab.~\ref{tab:3b-mudependence-data}. It can be observed that the scale dependence is obviously weaker at the NLO than at the LO. Nevertheless, at the NLO, the uncertainty for the $\Omega_{bbb}^{-}$ mass due to the scale dependence is still not small. This implies that all estimates of the masses at the LO in the QCD sum rules must suffer from large uncertainties caused by the scale dependence, and can be improved by including the NLO contributions.

Our result for the $\Omega_{bbb}^{-}$ mass at the LO ($13.97^{+0.54}_{-0.50}$ GeV) is comparable with previous works in QCD sum rules at the LO~\cite{Zhang:2009re,Wang:2020avt} ($13.28 \pm0.10~{\rm{GeV}}$ and $14.43 \pm0.09~{\rm{GeV}}$  respectively), which is similar to the case of $\Omega_{ccc}^{++}$. Theoretical uncertainties are significantly reduced after considering the NLO contribution. We then obtain $14.27^{+0.33}_{-0.32} $ GeV at the NLO in the $\overline{\rm{MS}}$ scheme. The results obtained in various approaches are listed in Tab.~\ref{tab:3c-3b-DiffModels}, all of which are comparable with our results.

\begin{table}[H]
  \renewcommand\arraystretch{2}
  \begin{center}
  \setlength{\tabcolsep}{4 mm}
   \caption{Predicted masses of $\Omega_{ccc}$ and $\Omega_{bbb}$ in different approaches. Here the central values of our results are obtained with the renormalization scale $\mu=M_B$. (With $\mu=(1.2-2.0)M_B$ the $\Omega_{ccc}$ mass is calculated as $4.75-4.80~GeV$, see Tab.2 and Fig.4.$)^*$}
\begin{tabular}{lc|@{\ }|c}
 \hline\hline
    Models    &     $\Omega_{ccc}\ (GeV)$ & $\Omega_{bbb}\ (GeV)$\\ \hline

    Lattic QCD \cite{Briceno:2012wt,Meinel:2012qz,Padmanath:2013zfa,Namekawa:2013vu,Brown:2014ena,Alexandrou:2014sha,Can:2015exa}   &     4.7 $ \sim$ 4.8    &   14.36 $ \sim$ 14.37 \\
                                                                      QCD Sum Rule \cite{Zhang:2009re,Aliev:2014lxa,Wang:2020avt}   &     4.6 $ \sim$ 5.0    &   13.28 $ \sim$ 14.83  \\
    Various potential models \cite{Hasenfratz:1980ka,Vijande:2004at,Migura:2006ep,Roberts:2007ni,Martynenko:2007je,Bernotas:2008bu,Flynn:2011gf,Vijande:2015faa,Thakkar:2016sog,Chen:2016spr,Weng:2018mmf,Yang:2019lsg,Liu:2019vtx}   &     4.76 $ \sim$ 4.90    &   14.27 $ \sim$ 14.83   \\
    Fadeev equation \cite{Radin:2014yna,Qin:2019hgk,Yin:2019bxe,Gutierrez-Guerrero:2019uwa}                                          &     4.76 $ \sim$ 5.00    &   14.23 $ \sim$ 14.57   \\
    Regge trajectories \cite{Wei:2015gsa,Wei:2016jyk}                                                                           &     4.834     &         14.788  \\
    This work                                                                          &  \makecell{$4.53^{+0.26}_{-0.11}$ \\$4.75 - 4.80^*$}     &         $14.27^{+0.33}_{-0.32}$  \\ \hline\hline
   \end{tabular}
   \label{tab:3c-3b-DiffModels}
   \end{center}
   \end{table}

\subsection{Discussion of GGG Condensate Contributions}

In this work, we also evaluate the GGG condensate contributions, and find that are very samll. Thus, they do not affect the phenomenological results.

\begin{figure}[H]
\centering
\includegraphics[scale=1]{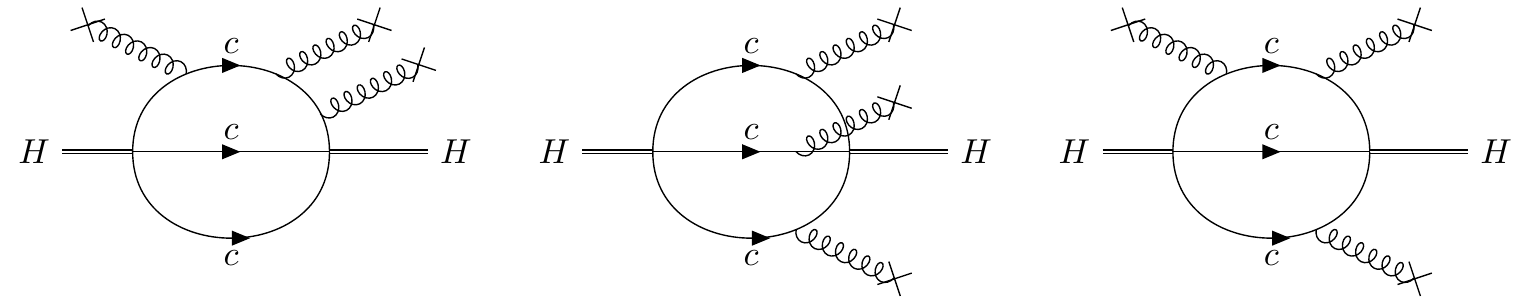}
\caption{\label{fig:GGG}
  Part contribution of $C_{GGG}$-LO. $H$ denotes the interpolating current.}
\end{figure}

The leading order contributions of the GGG condensate come from two parts: one is from the diagrams shown in Fig.\ref{fig:GGG} and the other is from the higher order of the gluons background field in the $C_{GG}$-LO of Fig.\ref{fig:FeynmanDiagrams-LO}, i.e., $\langle GGG\rangle$ that is related to the $\langle A(x)A(y)\rangle$ expansion. (Similar to the results of $C_1$ and $C_{GG}$, the results of $C_{GGG}$ will be shared by auxiliary files). Choosing $\langle g_s^3 \hat{G}\hat{G}\hat{G} \rangle\ =\ 0.054 \pm 0.014 \rm{GeV^6}$~\cite{Shifman:1978by,Reinders:1984sr,Colangelo:2000dp,Wang:2020avt}, similar to \cite{Wang:2020avt}, we find the ratios of different condensate contributions near the Borel platform as

\begin{align}
\int_{s_{\rm{th}}}^{s_0} ds \  \rho_{1}(s)\ e^{-\frac{s}{M_B^2}}  : \ \int_{s_{\rm{th}}}^\infty ds \   \rho_{GG}(s)  \langle g_s^2 \hat{G}\hat{G} \rangle \ e^{-\frac{s}{M_B^2}}\ : \ \ \int_{s_{\rm{th}}}^\infty ds \   \rho_{GGG}(s)  \langle g_s^3 \hat{G}\hat{G}\hat{G} \rangle \ e^{-\frac{s}{M_B^2}}\   \simeq \ 1\ : \mathcal{O}(10^{-2}) : \mathcal{O}(10^{-4})
\end{align}
Even if we use a larger value of $\langle g_s^3 \hat{G}\hat{G}\hat{G} \rangle\ =\ 0.52 \pm 0.10 \rm{GeV^6}$ as chosen in \cite{Albuquerque:2020hio,Narison:2010cg,Narison:2011xe,Narison:2011rn}, we can still observe that this d=6 GGG condensate contributions are still too small to affect the phenomenological results. This conclusion is consistent with the finding in Ref.\cite{Wang:2020avt}. Therefore, according to our study, the GGG condensate contributions can be ignored.


\section{Summary}

In the study of hadron physics, the QCD sum rules is known to be a powerful tool to evaluate hadron properties. For the triply heavy baryons $QQQ ~(Q=c,b)$, previous works only dealt with the LO perturbative QCD calculation. The absence of the higher order QCD corrections may lead to large theoretical uncertainties in QCD sum rules. In this paper, for the first time, we calculate the NLO QCD contribution of the $QQQ$ system by considering the NLO correction to the perturbative term of $C_1$ in the OPE. Additionally, the nonperturbative contribution is embodied by the d=4 gluon-gluon condensate. We also introduce adequate constraints to ensure the ground state dominance and power suppressions within the Borel sum rules.

Because the main theoretical uncertainty of the LO result is from the absence of the NLO correction of $C_1$ in QCD sum rules, the inclusion of the NLO contribution plays important roles theoretically and phenomenologically.
By comparing the NLO result with the LO one, we may draw the following conclusions for the $\Omega_{QQQ}$ system. First, the correction from the NLO contribution is sizable ($\Delta M_H =
M_H^{NLO}\, -\, M_H^{LO}$ can be as large as $ 0.2 \sim 0.3\,  GeV$ or even larger).
Second, after considering the NLO contribution, the parameters dependence of the results is reduced and the Borel platform becomes more distinct, especially for the $bbb$ system in the $\overline{\rm{MS}}$ scheme.
Third, the dependences on the renormalization schemes and quark masses are significantly improved for the $\Omega_{ccc}$.

By including the NLO contribution of the perturbative part in QCD sum rules, we find the masses to be $4.53^{+0.26}_{-0.11}$ GeV for $\Omega_{ccc}$ and $14.27^{+0.33}_{-0.32}$  GeV for $\Omega_{bbb}$, where the results are obtained at $\mu=M_B$ with errors including that from the variation of the renormalization scale $\mu$ in the range $(0.8-1.2) M_B$.
Further study for the $\mu$ dependence in a wider range shows that the LO results are very sensitive to the choice of $\mu$, and therefore have large uncertainties, whereas the NLO results are considerably better. Aside from the masses given above at $\mu=M_B$ (in Tab.~\ref{tab:3c-3b-DiffModels}), a quite stable value, (4.75-4.80) GeV, for the $\Omega_{ccc}$ mass is found in the range of $\mu=(1.2-2.0) M_B$. The distinctions between the LO and NLO results on the renormalization scale can be most clearly observed in  Fig.[\ref{fig:3c-mu--kMB}] and  Tab.~\ref{tab:3c-mudependence-data} for the $\Omega_{ccc}$. Considering all important factors, like $\mu$ dependence and perturbative convergence, 4.75-4.80 GeV is a more credible prediction for the $\Omega_{ccc}$ mass. Finally, we also investigate d=6 GGG condensate contributions, which are found to be so small that they can be ignored in the phenomenological analysis.

Therefore, in our view, the NLO contribution is indeed important and should not be ignored. Further investigations on the NLO corrections, e.g., to the coefficient  $C_{GG}$ of the two gluon condensation, may be needed, which will be helpful to constrain the theoretical errors for the nonperturbative contributions.

\section*{acknowledgments}
We thank Chen-Yu Wang, Xiao Liu, and Xin Guan for the many useful and helpful discussions. We also thank Shi-Lin Zhu for the helpful comments. The work is supported in part by the National Natural Science Foundation of China (Grants No. 11875071, No. 11975029) and the National Key Research and Development Program of China under Contracts No. 2020YFA0406400.

\newpage
\appendix

\section{Calculation Of Operator Renormalization Term}

The definition of the baryon operator is shown as follows,
\begin{align}
\mathcal{O}_{\Gamma_1,\Gamma_2} = \epsilon_{ijk}\left( (q_1^i)^T \hat{C}\Gamma_1 q_2^j \right)\Gamma_2 q_3^k\, .
\end{align}
and the next-to-leading order correction excluding the quark self-energy diagrams, can be divided into two parts,
\begin{align}
A=\int \frac{d_p^D}{(2\pi)^D}(A_1+A_2)\ ,
\end{align}
where $A_1$ denotes the case of the exchanging gluon between quarks $q_1$ and $q_2$, and $A_2$ denotes the other cases of the exchanging gluon between quarks $q_1$ and $q_3$, $q_2$ and $q_3$. The corresponding amplitudes are
\begin{align}
A_{1} & =\epsilon_{ijk} \left[i g_s \gamma_{\mu}  \frac{-i\slashed{p}}{p^{2}}\Gamma_{1}  \frac{i\slashed{p}}{p^{2}} i g_s \gamma^{\mu} \left(T^{a}\right)_{i i^{\prime}}\left(T^{a}\right)_{j j^{\prime}}\right] \left[\Gamma_{2} \delta_{k k^{\prime}} \right]  \frac{-i}{p^{2}}\ ,
\end{align}

\begin{align}
A_{2} & = \epsilon_{ijk}\left[i g_s \gamma_{\mu}  \frac{-i\slashed{p}}{p^{2}} \Gamma_{1} \left(T^{a}\right)_{i i^{\prime}} \delta_{j j^{\prime}} +\Gamma_{1}  \frac{-i\slashed{p}}{p^{2}} i g_s \gamma_{\mu} \delta_{i i^{\prime}}\left(T^{a}\right)_{j j^{\prime}} \right] \left[\Gamma_{2} i g_s \gamma^{\mu} \frac{i\slashed{p}}{p^{2}} \left(T^{a}\right)_{k k^{\prime}} \right] \frac{-i}{p^{2}}\ ,
\end{align}
Because there are no residual infrared(IR) divergences, we only need to consider ultraviolet(UV) divergences. Therefore the mass terms are ignored in quark propagators.

Then we can get the following form after the reduction:
\begin{align}\label{eq:A}
A=-i g_s^2 \frac{1}{D} \int \frac{d_p^D}{(2\pi)^D}\ \frac{1}{(p^2)^2}\ B\ ,
\end{align}
where
\begin{align}
\begin{split}
B &=-\left(\gamma_{\mu} \gamma_{\nu} \Gamma_{1} \gamma^{\nu} \gamma^{\mu}\right)\left(\Gamma_{2}\right)\left(T^{a}\right)_{i i^{\prime}}\left(T^{a}\right)_{j j^{\prime}} \delta_{k k^{\prime}}\epsilon_{ijk} -D \left(\Gamma_{1}\right)\left(\Gamma_{2}\right)\left[\left(T^{a}\right)_{i i^{\prime}} \delta_{j j^{\prime}}+\delta_{i i^{\prime}}\left(T^{a}\right)_{j j^{\prime}}\right]\left(T^{a}\right)_{k k} \epsilon_{ijk}\\
&-\frac{1}{2}\left(\left[\sigma_{\mu \nu}, \Gamma_{1}\right]\right)\left(\Gamma_{2} \sigma_{\mu \nu}\right)\left[\left(T^{a}\right)_{i i^{\prime}} \delta_{j j^{\prime}}+\delta_{i i^{\prime}}\left(T^{a}\right)_{j j^{\prime}}\right]\left(T^{a}\right)_{k k^{\prime}}\epsilon_{ijk}\ .
\end{split}
\end{align}

According to Eq.[\ref{eq:A}], we can get the UV divergences part of A.
\begin{align}
A_{UV} & = -\left.i g_s^{2} \frac{1}{4} \frac{i}{(4 \pi)^{2}} \frac{1}{\varepsilon} B\right|_{D  = 4}  = \frac{\alpha_{s}}{\varepsilon} \frac{\left.B\right|_{D  = 4}}{16\pi}\ ,
\end{align}
$A_{UV}$ is canceled by the operator renormalization term and renormalization coefficients of the quark wave function from operator $\mathcal{O}_{\Gamma_1,\Gamma_2}$, thus, the corresponding operator renormalization term in $\overline{\rm{MS}}$ scheme is
\begin{align}
Z_O =1+\delta Z_O=1+ \delta \left( \frac{1}{\epsilon}+Log(4\pi)-\gamma_E \right) \ ,
\end{align}
where
\begin{align}
\delta=- \alpha_s \frac{\left.B\right|_{D  = 4}}{16\pi}- 3\left(-\frac{1}{2} \frac{\alpha_s}{3\pi} \right) \textbf{I}\ .
\end{align}
where $\textbf{I}$ denotes the identity matrix.

At last, in this work, we can use the following relation, gotten by the Fierz transformation.
 \begin{align}
\epsilon_{ijk}\left( (Q^i)^T \hat{C}\gamma_\lambda Q^j \right) \sigma_{\lambda \mu} Q^k\ = \ i\ \epsilon_{ijk}\left( (Q^i)^T \hat{C}\gamma_\mu Q^j \right) Q^k.
\end{align}
This relation is valid in the D dimension, ($D=4-2\epsilon$).

\newpage

\begin{thebibliography}{10}

\bibitem{Tanabashi:2018oca}
{\bfseries Particle Data Group} , M.~Tanabashi {\em et al.}, {\it {Review of
  Particle Physics}},  \href{http://dx.doi.org/10.1103/PhysRevD.98.030001}{{\em
  Phys. Rev. D} {\bfseries 98} (2018) 030001}.

\bibitem{Chen:2016qju}
H.-X. Chen, W.~Chen, X.~Liu, and S.-L. Zhu, {\it {The hidden-charm pentaquark
  and tetraquark states}},
  \href{http://dx.doi.org/10.1016/j.physrep.2016.05.004}{{\em Phys. Rept.}
  {\bfseries 639} (2016) 1--121}
  [\href{http://arxiv.org/abs/1601.02092}{{\ttfamily arXiv:1601.02092}}].

\bibitem{Liu:2019zoy}
Y.-R. Liu, H.-X. Chen, W.~Chen, X.~Liu, and S.-L. Zhu, {\it {Pentaquark and
  Tetraquark states}},
  \href{http://dx.doi.org/10.1016/j.ppnp.2019.04.003}{{\em Prog. Part. Nucl.
  Phys.} {\bfseries 107} (2019) 237--320}
  [\href{http://arxiv.org/abs/1903.11976}{{\ttfamily arXiv:1903.11976}}].

\bibitem{Brambilla:2019esw}
N.~Brambilla, S.~Eidelman, C.~Hanhart, A.~Nefediev, C.-P. Shen, C.~E. Thomas,
  A.~Vairo, and C.-Z. Yuan, {\it {The $XYZ$ states: experimental and
  theoretical status and perspectives}},
  \href{http://dx.doi.org/10.1016/j.physrep.2020.05.001}{{\em Phys. Rept.}
  {\bfseries 873} (2020) 1--154}
  [\href{http://arxiv.org/abs/1907.07583}{{\ttfamily arXiv:1907.07583}}].

\bibitem{Aaij:2017ueg}
{\bfseries LHCb} , R.~Aaij {\em et al.}, {\it {Observation of the doubly
  charmed baryon $\Xi_{cc}^{++}$}},
  \href{http://dx.doi.org/10.1103/PhysRevLett.119.112001}{{\em Phys. Rev.
  Lett.} {\bfseries 119} (2017) 112001}
  [\href{http://arxiv.org/abs/1707.01621}{{\ttfamily arXiv:1707.01621}}].

\bibitem{Saleev:1999ti}
V.~A. Saleev, {\it {Omega(ccc) production via fragmentation at LHC}},
  \href{http://dx.doi.org/10.1142/S0217732399002741}{{\em Mod. Phys. Lett. A}
  {\bfseries 14} (1999) 2615--2620}
  [\href{http://arxiv.org/abs/hep-ph/9906515}{{\ttfamily hep-ph/9906515}}].

\bibitem{GomshiNobary:2003sf}
M.~A. Gomshi~Nobary, {\it {Fragmentation production of Omega(ccc) and
  Omega(bbb) baryons}},
  \href{http://dx.doi.org/10.1016/j.physletb.2002.12.001}{{\em Phys. Lett. B}
  {\bfseries 559} (2003) 239--244}
  [\href{http://arxiv.org/abs/hep-ph/0408122}{{\ttfamily hep-ph/0408122}}].
  [Erratum: Phys.Lett.B 598, 294--294 (2004)].

\bibitem{GomshiNobary:2004mq}
M.~A. Gomshi~Nobary and R.~Sepahvand, {\it {Fragmentation of triply heavy
  baryons}},  \href{http://dx.doi.org/10.1103/PhysRevD.71.034024}{{\em Phys.
  Rev. D} {\bfseries 71} (2005) 034024}
  [\href{http://arxiv.org/abs/hep-ph/0406148}{{\ttfamily hep-ph/0406148}}].

\bibitem{GomshiNobary:2005ur}
M.~A. Gomshi~Nobary and R.~Sepahvand, {\it {An Ivestigation of triply heavy
  baryon production at hadron colliders}},
  \href{http://dx.doi.org/10.1016/j.nuclphysb.2006.01.043}{{\em Nucl. Phys. B}
  {\bfseries 741} (2006) 34--41}
  [\href{http://arxiv.org/abs/hep-ph/0508115}{{\ttfamily hep-ph/0508115}}].

\bibitem{Chen:2011mb}
Y.-Q. Chen and S.-Z. Wu, {\it {Production of Triply Heavy Baryons at LHC}},
  \href{http://dx.doi.org/10.1007/JHEP08(2011)144}{{\em JHEP} {\bfseries 08}
  (2011) 144} [\href{http://arxiv.org/abs/1106.0193}{{\ttfamily
  arXiv:1106.0193}}]. [Erratum: JHEP 09, 089 (2011)].

\bibitem{He:2014tga}
H.~He, Y.~Liu, and P.~Zhuang, {\it {$\Omega_{ccc}$ production in high energy
  nuclear collisions}},
  \href{http://dx.doi.org/10.1016/j.physletb.2015.04.049}{{\em Phys. Lett. B}
  {\bfseries 746} (2015) 59--63}
  [\href{http://arxiv.org/abs/1409.1009}{{\ttfamily arXiv:1409.1009}}].

\bibitem{Baranov:2004er}
S.~P. Baranov and V.~L. Slad, {\it {Production of triply charmed Omega(ccc)
  baryons in e+ e- annihilation}},
  \href{http://dx.doi.org/10.1134/1.1707141}{{\em Phys. Atom. Nucl.} {\bfseries
  67} (2004) 808--814} [\href{http://arxiv.org/abs/hep-ph/0603090}{{\ttfamily
  hep-ph/0603090}}].

\bibitem{Bjorken:1985ei}
J.~D. Bjorken, {\it {Is the ccc a new deal for baryon spectroscopy?}},
  \href{http://dx.doi.org/10.1063/1.35379}{{\em AIP Conf. Proc.} {\bfseries
  132} (1985) 390--403}.

\bibitem{Flynn:2011gf}
J.~M. Flynn, E.~Hernandez, and J.~Nieves, {\it {Triply Heavy Baryons and Heavy
  Quark Spin Symmetry}},
  \href{http://dx.doi.org/10.1103/PhysRevD.85.014012}{{\em Phys. Rev. D}
  {\bfseries 85} (2012) 014012}
  [\href{http://arxiv.org/abs/1110.2962}{{\ttfamily arXiv:1110.2962}}].

\bibitem{Wang:2018utj}
W.~Wang and J.~Xu, {\it {Weak Decays of Triply Heavy Baryons}},
  \href{http://dx.doi.org/10.1103/PhysRevD.97.093007}{{\em Phys. Rev. D}
  {\bfseries 97} (2018) 093007}
  [\href{http://arxiv.org/abs/1803.01476}{{\ttfamily arXiv:1803.01476}}].

\bibitem{Briceno:2012wt}
R.~A. Briceno, H.-W. Lin, and D.~R. Bolton, {\it {Charmed-Baryon Spectroscopy
  from Lattice QCD with $N_f$ = 2+1+1 Flavors}},
  \href{http://dx.doi.org/10.1103/PhysRevD.86.094504}{{\em Phys. Rev. D}
  {\bfseries 86} (2012) 094504}
  [\href{http://arxiv.org/abs/1207.3536}{{\ttfamily arXiv:1207.3536}}].

\bibitem{Meinel:2012qz}
S.~Meinel, {\it {Excited-state spectroscopy of triply-bottom baryons from
  lattice QCD}},  \href{http://dx.doi.org/10.1103/PhysRevD.85.114510}{{\em
  Phys. Rev. D} {\bfseries 85} (2012) 114510}
  [\href{http://arxiv.org/abs/1202.1312}{{\ttfamily arXiv:1202.1312}}].

\bibitem{Padmanath:2013zfa}
M.~Padmanath, R.~G. Edwards, N.~Mathur, and M.~Peardon, {\it {Spectroscopy of
  triply-charmed baryons from lattice QCD}},
  \href{http://dx.doi.org/10.1103/PhysRevD.90.074504}{{\em Phys. Rev. D}
  {\bfseries 90} (2014) 074504}
  [\href{http://arxiv.org/abs/1307.7022}{{\ttfamily arXiv:1307.7022}}].

\bibitem{Namekawa:2013vu}
{\bfseries PACS-CS} , Y.~Namekawa {\em et al.}, {\it {Charmed baryons at the
  physical point in 2+1 flavor lattice QCD}},
  \href{http://dx.doi.org/10.1103/PhysRevD.87.094512}{{\em Phys. Rev. D}
  {\bfseries 87} (2013) 094512}
  [\href{http://arxiv.org/abs/1301.4743}{{\ttfamily arXiv:1301.4743}}].

\bibitem{Brown:2014ena}
Z.~S. Brown, W.~Detmold, S.~Meinel, and K.~Orginos, {\it {Charmed bottom baryon
  spectroscopy from lattice QCD}},
  \href{http://dx.doi.org/10.1103/PhysRevD.90.094507}{{\em Phys. Rev. D}
  {\bfseries 90} (2014) 094507}
  [\href{http://arxiv.org/abs/1409.0497}{{\ttfamily arXiv:1409.0497}}].

\bibitem{Alexandrou:2014sha}
C.~Alexandrou, V.~Drach, K.~Jansen, C.~Kallidonis, and G.~Koutsou, {\it {Baryon
  spectrum with $N_f=2+1+1$ twisted mass fermions}},
  \href{http://dx.doi.org/10.1103/PhysRevD.90.074501}{{\em Phys. Rev. D}
  {\bfseries 90} (2014) 074501}
  [\href{http://arxiv.org/abs/1406.4310}{{\ttfamily arXiv:1406.4310}}].

\bibitem{Can:2015exa}
K.~U. Can, G.~Erkol, M.~Oka, and T.~T. Takahashi, {\it {Look inside
  charmed-strange baryons from lattice QCD}},
  \href{http://dx.doi.org/10.1103/PhysRevD.92.114515}{{\em Phys. Rev. D}
  {\bfseries 92} (2015) 114515}
  [\href{http://arxiv.org/abs/1508.03048}{{\ttfamily arXiv:1508.03048}}].

\bibitem{Zhang:2009re}
J.-R. Zhang and M.-Q. Huang, {\it {Deciphering triply heavy baryons in terms of
  QCD sum rules}},
  \href{http://dx.doi.org/10.1016/j.physletb.2009.02.056}{{\em Phys. Lett. B}
  {\bfseries 674} (2009) 28--35}
  [\href{http://arxiv.org/abs/0902.3297}{{\ttfamily arXiv:0902.3297}}].

\bibitem{Aliev:2014lxa}
T.~M. Aliev, K.~Azizi, and M.~Savc\i{}, {\it {Properties of triply heavy
  spin-3/2 baryons}},
  \href{http://dx.doi.org/10.1088/0954-3899/41/6/065003}{{\em J. Phys. G}
  {\bfseries 41} (2014) 065003}
  [\href{http://arxiv.org/abs/1404.2091}{{\ttfamily arXiv:1404.2091}}].

\bibitem{Wang:2020avt}
Z.-G. Wang, {\it {Analysis of the triply-heavy baryon states with the QCD sum
  rules}},  [\href{http://arxiv.org/abs/2010.08939}{{\ttfamily
  arXiv:2010.08939}}].

\bibitem{Hasenfratz:1980ka}
P.~Hasenfratz, R.~R. Horgan, J.~Kuti, and J.~M. Richard, {\it {Heavy Baryon
  Spectroscopy in the \{QCD\} Bag Model}},
  \href{http://dx.doi.org/10.1016/0370-2693(80)90906-5}{{\em Phys. Lett. B}
  {\bfseries 94} (1980) 401--404}.

\bibitem{Vijande:2004at}
J.~Vijande, H.~Garcilazo, A.~Valcarce, and F.~Fernandez, {\it {Spectroscopy of
  doubly charmed baryons}},
  \href{http://dx.doi.org/10.1103/PhysRevD.70.054022}{{\em Phys. Rev. D}
  {\bfseries 70} (2004) 054022}
  [\href{http://arxiv.org/abs/hep-ph/0408274}{{\ttfamily hep-ph/0408274}}].

\bibitem{Migura:2006ep}
S.~Migura, D.~Merten, B.~Metsch, and H.-R. Petry, {\it {Charmed baryons in a
  relativistic quark model}},
  \href{http://dx.doi.org/10.1140/epja/i2006-10017-9}{{\em Eur. Phys. J. A}
  {\bfseries 28} (2006) 41}
  [\href{http://arxiv.org/abs/hep-ph/0602153}{{\ttfamily hep-ph/0602153}}].

\bibitem{Roberts:2007ni}
W.~Roberts and M.~Pervin, {\it {Heavy baryons in a quark model}},
  \href{http://dx.doi.org/10.1142/S0217751X08041219}{{\em Int. J. Mod. Phys. A}
  {\bfseries 23} (2008) 2817--2860}
  [\href{http://arxiv.org/abs/0711.2492}{{\ttfamily arXiv:0711.2492}}].

\bibitem{Martynenko:2007je}
A.~P. Martynenko, {\it {Ground-state triply and doubly heavy baryons in a
  relativistic three-quark model}},
  \href{http://dx.doi.org/10.1016/j.physletb.2008.04.030}{{\em Phys. Lett. B}
  {\bfseries 663} (2008) 317--321}
  [\href{http://arxiv.org/abs/0708.2033}{{\ttfamily arXiv:0708.2033}}].

\bibitem{Bernotas:2008bu}
A.~Bernotas and V.~Simonis, {\it {Heavy hadron spectroscopy and the bag
  model}},  \href{http://dx.doi.org/10.3952/lithjphys.49110}{{\em Lith. J.
  Phys.} {\bfseries 49} (2009) 19--28}
  [\href{http://arxiv.org/abs/0808.1220}{{\ttfamily arXiv:0808.1220}}].

\bibitem{Vijande:2015faa}
J.~Vijande, A.~Valcarce, and H.~Garcilazo, {\it {Constituent-quark model
  description of triply heavy baryon nonperturbative lattice QCD data}},
  \href{http://dx.doi.org/10.1103/PhysRevD.91.054011}{{\em Phys. Rev. D}
  {\bfseries 91} (2015) 054011}
  [\href{http://arxiv.org/abs/1507.03735}{{\ttfamily arXiv:1507.03735}}].

\bibitem{Thakkar:2016sog}
K.~Thakkar, A.~Majethiya, and P.~C. Vinodkumar, {\it {Magnetic moments of
  baryons containing all heavy quarks in the quark-diquark model}},
  \href{http://dx.doi.org/10.1140/epjp/i2016-16339-4}{{\em Eur. Phys. J. Plus}
  {\bfseries 131} (2016) 339}
  [\href{http://arxiv.org/abs/1609.05444}{{\ttfamily arXiv:1609.05444}}].

\bibitem{Chen:2016spr}
H.-X. Chen, W.~Chen, X.~Liu, Y.-R. Liu, and S.-L. Zhu, {\it {A review of the
  open charm and open bottom systems}},
  \href{http://dx.doi.org/10.1088/1361-6633/aa6420}{{\em Rept. Prog. Phys.}
  {\bfseries 80} (2017) 076201}
  [\href{http://arxiv.org/abs/1609.08928}{{\ttfamily arXiv:1609.08928}}].

\bibitem{Weng:2018mmf}
X.-Z. Weng, X.-L. Chen, and W.-Z. Deng, {\it {Masses of doubly heavy-quark
  baryons in an extended chromomagnetic model}},
  \href{http://dx.doi.org/10.1103/PhysRevD.97.054008}{{\em Phys. Rev. D}
  {\bfseries 97} (2018) 054008}
  [\href{http://arxiv.org/abs/1801.08644}{{\ttfamily arXiv:1801.08644}}].

\bibitem{Yang:2019lsg}
G.~Yang, J.~Ping, P.~G. Ortega, and J.~Segovia, {\it {Triply heavy baryons in
  the constituent quark model}},
  \href{http://dx.doi.org/10.1088/1674-1137/44/2/023102}{{\em Chin. Phys. C}
  {\bfseries 44} (2020) 023102}
  [\href{http://arxiv.org/abs/1904.10166}{{\ttfamily arXiv:1904.10166}}].

\bibitem{Liu:2019vtx}
M.-S. Liu, Q.-F. L\"u, and X.-H. Zhong, {\it {Triply charmed and bottom baryons
  in a constituent quark model}},
  \href{http://dx.doi.org/10.1103/PhysRevD.101.074031}{{\em Phys. Rev. D}
  {\bfseries 101} (2020) 074031}
  [\href{http://arxiv.org/abs/1912.11805}{{\ttfamily arXiv:1912.11805}}].

\bibitem{Radin:2014yna}
M.~Radin, S.~Babaghodrat, and M.~Monemzadeh, {\it {Estimation of heavy baryon
  masses \ensuremath{\Omega}ccc++ and \ensuremath{\Omega}bbb- by solving the
  Faddeev equation in a three-dimensional approach}},
  \href{http://dx.doi.org/10.1103/PhysRevD.90.047701}{{\em Phys. Rev. D}
  {\bfseries 90} (2014) 047701}.

\bibitem{Qin:2019hgk}
S.-x. Qin, C.~D. Roberts, and S.~M. Schmidt, {\it {Spectrum of light- and
  heavy-baryons}},  \href{http://dx.doi.org/10.1007/s00601-019-1488-x}{{\em Few
  Body Syst.} {\bfseries 60} (2019) 26}
  [\href{http://arxiv.org/abs/1902.00026}{{\ttfamily arXiv:1902.00026}}].

\bibitem{Yin:2019bxe}
P.-L. Yin, C.~Chen, G.~a. Krein, C.~D. Roberts, J.~Segovia, and S.-S. Xu, {\it
  {Masses of ground-state mesons and baryons, including those with heavy
  quarks}},  \href{http://dx.doi.org/10.1103/PhysRevD.100.034008}{{\em Phys.
  Rev. D} {\bfseries 100} (2019) 034008}
  [\href{http://arxiv.org/abs/1903.00160}{{\ttfamily arXiv:1903.00160}}].

\bibitem{Gutierrez-Guerrero:2019uwa}
L.~X. Guti\'errez-Guerrero, A.~Bashir, M.~A. Bedolla, and E.~Santopinto, {\it
  {Masses of Light and Heavy Mesons and Baryons: A Unified Picture}},
  \href{http://dx.doi.org/10.1103/PhysRevD.100.114032}{{\em Phys. Rev. D}
  {\bfseries 100} (2019) 114032}
  [\href{http://arxiv.org/abs/1911.09213}{{\ttfamily arXiv:1911.09213}}].

\bibitem{Wei:2015gsa}
K.-W. Wei, B.~Chen, and X.-H. Guo, {\it {Masses of doubly and triply charmed
  baryons}},  \href{http://dx.doi.org/10.1103/PhysRevD.92.076008}{{\em Phys.
  Rev. D} {\bfseries 92} (2015) 076008}
  [\href{http://arxiv.org/abs/1503.05184}{{\ttfamily arXiv:1503.05184}}].

\bibitem{Wei:2016jyk}
K.-W. Wei, B.~Chen, N.~Liu, Q.-Q. Wang, and X.-H. Guo, {\it {Spectroscopy of
  singly, doubly, and triply bottom baryons}},
  \href{http://dx.doi.org/10.1103/PhysRevD.95.116005}{{\em Phys. Rev. D}
  {\bfseries 95} (2017) 116005}
  [\href{http://arxiv.org/abs/1609.02512}{{\ttfamily arXiv:1609.02512}}].

\bibitem{Shifman:1978bx}
M.~A. Shifman, A.~Vainshtein, and V.~I. Zakharov, {\it {QCD and Resonance
  Physics. Theoretical Foundations}},
  \href{http://dx.doi.org/10.1016/0550-3213(79)90022-1}{{\em Nucl. Phys. B}
  {\bfseries 147} (1979) 385--447}.

\bibitem{Shifman:1978by}
M.~A. Shifman, A.~I. Vainshtein, and V.~I. Zakharov, {\it {QCD and Resonance
  Physics: Applications}},
  \href{http://dx.doi.org/10.1016/0550-3213(79)90023-3}{{\em Nucl. Phys. B}
  {\bfseries 147} (1979) 448--518}.

\bibitem{Reinders:1984sr}
L.~Reinders, H.~Rubinstein, and S.~Yazaki, {\it {Hadron Properties from QCD Sum
  Rules}},  \href{http://dx.doi.org/10.1016/0370-1573(85)90065-1}{{\em Phys.
  Rept.} {\bfseries 127} (1985) 1}.

\bibitem{Colangelo:2000dp}
P.~Colangelo and A.~Khodjamirian, {\it {QCD sum rules, a modern perspective}},
  [\href{http://arxiv.org/abs/hep-ph/0010175}{{\ttfamily hep-ph/0010175}}].

\bibitem{Narison:2010wb}
S.~Narison, {\it {SVZ sum rules : 30 + 1 years later}},
  \href{http://dx.doi.org/10.1016/j.nuclphysbps.2010.10.078}{{\em Nucl. Phys. B
  Proc. Suppl.} {\bfseries 207-208} (2010) 315--322}
  [\href{http://arxiv.org/abs/1010.1959}{{\ttfamily arXiv:1010.1959}}].

\bibitem{Narison:2014wqa}
S.~Narison, {\it {Mini-review on QCD spectral sum rules}},
  \href{http://dx.doi.org/10.1016/j.nuclphysbps.2015.01.041}{{\em Nucl. Part.
  Phys. Proc.} {\bfseries 258-259} (2015) 189--194}
  [\href{http://arxiv.org/abs/1409.8148}{{\ttfamily arXiv:1409.8148}}].

\bibitem{Albuquerque:2018jkn}
R.~M. Albuquerque, J.~M. Dias, K.~P. Khemchandani, A.~Mart\'\i{}nez~Torres,
  F.~S. Navarra, M.~Nielsen, and C.~M. Zanetti, {\it {QCD sum rules approach to
  the $X,~Y$ and $Z$ states}},
  \href{http://dx.doi.org/10.1088/1361-6471/ab2678}{{\em J. Phys. G} {\bfseries
  46} (2019) 093002} [\href{http://arxiv.org/abs/1812.08207}{{\ttfamily
  arXiv:1812.08207}}].

\bibitem{Ovchinnikov:1991mu}
A.~A. Ovchinnikov, A.~A. Pivovarov, and L.~R. Surguladze, {\it {Baryonic sum
  rules in the next-to-leading order in alpha-s}},
  \href{http://dx.doi.org/10.1142/S0217751X91001015}{{\em Int. J. Mod. Phys. A}
  {\bfseries 6} (1991) 2025--2034}.

\bibitem{Groote:2008hz}
S.~Groote, J.~G. Korner, and A.~A. Pivovarov, {\it {Next-to-Leading Order
  perturbative QCD corrections to baryon correlators in matter}},
  \href{http://dx.doi.org/10.1103/PhysRevD.78.034039}{{\em Phys. Rev. D}
  {\bfseries 78} (2008) 034039}
  [\href{http://arxiv.org/abs/0805.3590}{{\ttfamily arXiv:0805.3590}}].

\bibitem{Groote:2008dx}
S.~Groote, J.~G. Korner, and A.~A. Pivovarov, {\it {Heavy baryon properties
  with NLO accuracy in perturbative QCD}},
  \href{http://dx.doi.org/10.1140/epjc/s10052-008-0763-7}{{\em Eur. Phys. J. C}
  {\bfseries 58} (2008) 355--382}
  [\href{http://arxiv.org/abs/0807.2148}{{\ttfamily arXiv:0807.2148}}].

\bibitem{Wang:2017qvg}
C.-Y. Wang, C.~Meng, Y.-Q. Ma, and K.-T. Chao, {\it {NLO effects for doubly
  heavy baryons in QCD sum rules}},
  \href{http://dx.doi.org/10.1103/PhysRevD.99.014018}{{\em Phys. Rev. D}
  {\bfseries 99} (2019) 014018}
  [\href{http://arxiv.org/abs/1708.04563}{{\ttfamily arXiv:1708.04563}}].

\bibitem{Ioffe:1981kw}
B.~L. Ioffe, {\it {Calculation of Baryon Masses in Quantum Chromodynamics}},
  \href{http://dx.doi.org/10.1016/0550-3213(81)90259-5}{{\em Nucl. Phys. B}
  {\bfseries 188} (1981) 317--341}. [Erratum: Nucl.Phys.B 191, 591--592
  (1981)].

\bibitem{Shifman:2000jv}
M.~A. Shifman, \href{http://dx.doi.org/10.1142/9789812810458_0032}{{\it {Quark
  hadron duality}}, } in {\em {8th International Symposium on Heavy Flavor
  Physics}}.
\newblock World Scientific, Singapore,
\newblock 7, 2000 [\href{http://arxiv.org/abs/hep-ph/0009131}{{\ttfamily
  hep-ph/0009131}}].

\bibitem{Shifman:2001qm}
M.~Shifman, {\it {Lectures on quark hadron duality}},
  \href{http://dx.doi.org/10.1007/s10582-002-0080-6}{{\em Czech. J. Phys.}
  {\bfseries 52} (2002) B102--B135}.

\bibitem{Narison:2007spa}
S.~Narison, {\em {QCD as a Theory of Hadrons}: {From Partons to Confinement}},
  vol.~17.
\newblock Cambridge University Press,
\newblock 7, 2007 [\href{http://arxiv.org/abs/hep-ph/0205006}{{\ttfamily
  hep-ph/0205006}}].

\bibitem{Kublbeck:1990xc}
J.~Kublbeck, M.~Bohm, and A.~Denner, {\it {Feyn Arts: Computer Algebraic
  Generation of Feynman Graphs and Amplitudes}},
  \href{http://dx.doi.org/10.1016/0010-4655(90)90001-H}{{\em Comput. Phys.
  Commun.} {\bfseries 60} (1990) 165--180}.

\bibitem{Hahn:2000kx}
T.~Hahn, {\it {Generating Feynman diagrams and amplitudes with FeynArts 3}},
  \href{http://dx.doi.org/10.1016/S0010-4655(01)00290-9}{{\em Comput. Phys.
  Commun.} {\bfseries 140} (2001) 418--431}
  [\href{http://arxiv.org/abs/hep-ph/0012260}{{\ttfamily hep-ph/0012260}}].

\bibitem{Mertig:1990an}
R.~Mertig, M.~Bohm, and A.~Denner, {\it {FEYN CALC: Computer algebraic
  calculation of Feynman amplitudes}},
  \href{http://dx.doi.org/10.1016/0010-4655(91)90130-D}{{\em Comput. Phys.
  Commun.} {\bfseries 64} (1991) 345--359}.

\bibitem{Shtabovenko:2016sxi}
V.~Shtabovenko, R.~Mertig, and F.~Orellana, {\it {New Developments in FeynCalc
  9.0}},  \href{http://dx.doi.org/10.1016/j.cpc.2016.06.008}{{\em Comput. Phys.
  Commun.} {\bfseries 207} (2016) 432--444}
  [\href{http://arxiv.org/abs/1601.01167}{{\ttfamily arXiv:1601.01167}}].

\bibitem{vonManteuffel:2012np}
A.~von Manteuffel and C.~Studerus, {\it {Reduze 2 - Distributed Feynman
  Integral Reduction}},  [\href{http://arxiv.org/abs/1201.4330}{{\ttfamily
  arXiv:1201.4330}}].

\bibitem{Kotikov:1990kg}
A.~V. Kotikov, {\it {Differential equations method: New technique for massive
  Feynman diagrams calculation}},
  \href{http://dx.doi.org/10.1016/0370-2693(91)90413-K}{{\em Phys. Lett. B}
  {\bfseries 254} (1991) 158--164}.

\bibitem{Bern:1992em}
Z.~Bern, L.~J. Dixon, and D.~A. Kosower, {\it {Dimensionally regulated one loop
  integrals}},  \href{http://dx.doi.org/10.1016/0370-2693(93)90400-C}{{\em
  Phys. Lett. B} {\bfseries 302} (1993) 299--308}
  [\href{http://arxiv.org/abs/hep-ph/9212308}{{\ttfamily hep-ph/9212308}}].
  [Erratum: Phys.Lett.B 318, 649 (1993)].

\bibitem{Remiddi:1997ny}
E.~Remiddi, {\it {Differential equations for Feynman graph amplitudes}},  {\em
  Nuovo Cim. A} {\bfseries 110} (1997) 1435--1452
  [\href{http://arxiv.org/abs/hep-th/9711188}{{\ttfamily hep-th/9711188}}].

\bibitem{Gehrmann:1999as}
T.~Gehrmann and E.~Remiddi, {\it {Differential equations for two loop four
  point functions}},
  \href{http://dx.doi.org/10.1016/S0550-3213(00)00223-6}{{\em Nucl. Phys. B}
  {\bfseries 580} (2000) 485--518}
  [\href{http://arxiv.org/abs/hep-ph/9912329}{{\ttfamily hep-ph/9912329}}].

\bibitem{Liu:2017jxz}
X.~Liu, Y.-Q. Ma, and C.-Y. Wang, {\it {A Systematic and Efficient Method to
  Compute Multi-loop Master Integrals}},
  \href{http://dx.doi.org/10.1016/j.physletb.2018.02.026}{{\em Phys. Lett. B}
  {\bfseries 779} (2018) 353--357}
  [\href{http://arxiv.org/abs/1711.09572}{{\ttfamily arXiv:1711.09572}}].

\bibitem{Bagan:1992za}
E.~Bagan, M.~Chabab, and S.~Narison, {\it {Baryons with two heavy quarks from
  QCD spectral sum rules}},
  \href{http://dx.doi.org/10.1016/0370-2693(93)90090-5}{{\em Phys. Lett. B}
  {\bfseries 306} (1993) 350--356}.

\bibitem{Dominguez:1994ce}
C.~Dominguez, G.~Gluckman, and N.~Paver, {\it {Mass of the charm quark from QCD
  sum rules}},  \href{http://dx.doi.org/10.1016/0370-2693(94)91027-8}{{\em
  Phys. Lett. B} {\bfseries 333} (1994) 184--189}
  [\href{http://arxiv.org/abs/hep-ph/9406329}{{\ttfamily hep-ph/9406329}}].

\bibitem{Dominguez:2014pga}
C.~Dominguez, L.~Hernandez, and K.~Schilcher, {\it {Determination of the gluon
  condensate from data in the charm-quark region}},
  \href{http://dx.doi.org/10.1007/JHEP07(2015)110}{{\em JHEP} {\bfseries 07}
  (2015) 110} [\href{http://arxiv.org/abs/1411.4500}{{\ttfamily
  arXiv:1411.4500}}].

\bibitem{Aoki:2016frl}
S.~Aoki {\em et al.}, {\it {Review of lattice results concerning low-energy
  particle physics}},
  \href{http://dx.doi.org/10.1140/epjc/s10052-016-4509-7}{{\em Eur. Phys. J. C}
  {\bfseries 77} (2017) 112} [\href{http://arxiv.org/abs/1607.00299}{{\ttfamily
  arXiv:1607.00299}}].

\bibitem{Zyla:2020zbs}
{\bfseries Particle Data Group} , P.~Zyla {\em et al.}, {\it {Review of
  Particle Physics}},  \href{http://dx.doi.org/10.1093/ptep/ptaa104}{{\em PTEP}
  {\bfseries 2020} (2020) 083C01}.

\bibitem{Bertlmann:1981he}
R.~A. Bertlmann, {\it {Heavy Quark - Anti-quark Systems From Exponential
  Moments in \{QCD\}}},
  \href{http://dx.doi.org/10.1016/0550-3213(82)90197-3}{{\em Nucl. Phys. B}
  {\bfseries 204} (1982) 387--412}.

\bibitem{Albuquerque:2020hio}
R.~M. Albuquerque, S.~Narison, A.~Rabemananjara, D.~Rabetiarivony, and
  G.~Randriamanatrika, {\it {Doubly-hidden scalar heavy molecules and
  tetraquarks states from QCD at NLO}},
  \href{http://dx.doi.org/10.1103/PhysRevD.102.094001}{{\em Phys. Rev. D}
  {\bfseries 102} (2020) 094001}
  [\href{http://arxiv.org/abs/2008.01569}{{\ttfamily arXiv:2008.01569}}].

\bibitem{Narison:2010cg}
S.~Narison, {\it {Gluon condensates and c, b quark masses from quarkonia ratios
  of moments}},  \href{http://dx.doi.org/10.1016/j.physletb.2011.09.116}{{\em
  Phys. Lett. B} {\bfseries 693} (2010) 559--566}
  [\href{http://arxiv.org/abs/1004.5333}{{\ttfamily arXiv:1004.5333}}].
  [Erratum: Phys.Lett.B 705, 544--544 (2011)].

\bibitem{Narison:2011xe}
S.~Narison, {\it {Gluon Condensates and precise $\overline{m}_{c,b}$ from
  QCD-Moments and their ratios to Order $\alpha_s^3$ and \ensuremath{<} G$^4$
  \ensuremath{>}}},
  \href{http://dx.doi.org/10.1016/j.physletb.2011.11.058}{{\em Phys. Lett. B}
  {\bfseries 706} (2012) 412--422}
  [\href{http://arxiv.org/abs/1105.2922}{{\ttfamily arXiv:1105.2922}}].

\bibitem{Narison:2011rn}
S.~Narison, {\it {Gluon Condensates and $\bar{m}_b(\bar{m}_b)$ from
  QCD-Exponential Moments at Higher Orders}},
  \href{http://dx.doi.org/10.1016/j.physletb.2011.12.047}{{\em Phys. Lett. B}
  {\bfseries 707} (2012) 259--263}
  [\href{http://arxiv.org/abs/1105.5070}{{\ttfamily arXiv:1105.5070}}].

\end{thebibliography}
\providecommand{\href}[2]{#2}\begingroup\raggedright\endgroup

\end{document}